\documentclass[preprint]{aastex}
\usepackage{epsf}
\usepackage{natbib}
\usepackage{subfigure}
\usepackage{epstopdf}
\usepackage{multirow,colortbl,longtable}

\newcommand{\be}{\begin{equation}}
\newcommand{\ee}{\end{equation}}

\newcommand{\feh}{\hbox{$ [{\rm Fe}/{\rm H}]$ }} 
\newcommand{\fehc}{\hbox{$ [{\rm Fe}/{\rm H}]$}}
\newcommand{\vmr}{$\hbox{\it V--R\/}$}
\def\c2{\chi ^2}

\begin{document}

\title{Period Change Similarities
among the RR Lyrae Variables in Oosterhoff I and Oosterhoff II Globular Systems\footnote{Based in part on observations made with the European Southern Observatory
telescopes obtained from the ESO/ST-ECF Science Archive Facility.}}

\author{Andrea Kunder and Alistair Walker} 
\affil{NOAO-Cerro Tololo Inter-American Observatory, Casilla 603, La Serena, Chile}
\affil{E-mail: akunder@ctio.noao.edu}

\author{Peter B. Stetson} 
\affil{Herzberg Institute of Astrophysics, National Research Council, Victoria, British Columbia V9E~2E7, Canada}

\author{Giuseppe Bono} 
\affil{Dipartimento di Fisica, Universita' di Roma Tor Vergata,
via della Ricerca Scientifica 1, 00133 Roma, Italy}
\affil{INAF-OAR, via Frascati 33, Monte Porzio Catone, Rome, Italy}

\author{James M. Nemec} 
\affil{Department of Physics \& Astronomy, Camosun College, Victoria, British Columbia, Canada}

\author{Roberto de Propris} 
\affil{NOAO-Cerro Tololo Inter-American Observatory, Casilla 603, La Serena, Chile}

\author{Matteo Monelli} 
\affil{IAC, Calle Via Lactea, E38200 La Laguna, Tenerife, Spain}

\author{Santi Cassisi} 
\affil{INAF-Osservatorio Astronomico di Collurania, 
via M. Maggini, 64100 Teramo, Italy}

\author{Gloria Andreuzzi} 
\affil{Fundaci\'{o}n Galileo Galilei - INAF, Bre\~{n}a Baja, Tenerife, Spain }

\author{Massimo Dall'Ora} 
\affil{INAF-Osservatorio Astronomico di Capodimonte, 
via Moiarello 16, 80131 Napoli, Italy}

\author{Alessandra Di Cecco} 
\affil{Dipartimento di Fisica, Universita' di Roma Tor Vergata, 
via della Ricerca Scientifica 1, 00133 Rome, Italy}
\affil{INAF-OAR, via Frascati 33, Monte Porzio Catone, Rome, Italy }

\author{Manuela Zoccali} 
\affil{Depto de Astronomia y Astrofisica, P. Universidad Catolica, Casilla 306,
Santiago 22, Chile}

\begin{abstract}
We present period change rates (dP/dt) for 42 RR Lyrae variables in the 
globular cluster IC$\,$4499.  Despite clear evidence of these period 
increases or decreases, the observed period change rates are an 
order of magnitude larger than predicted from theoretical models 
of this cluster.  We find there is a preference for increasing periods, 
a phenomenon observed in most  RR Lyrae stars in Milky Way globular
clusters.  The period-change rates as a function of
position in the period-amplitude plane are used to examine
possible evolutionary effects in OoI clusters, OoII clusters,
field RR Lyrae stars and the mixed-population cluster $\omega$~ Centauri.  
It is found that there is no correlation between the period
change rate and the typical definition of Oosterhoff groups.  
If the RR Lyrae period changes correspond with evolutionary effects,
this would be in contrast to the hypothesis that RR Lyrae 
variables in OoII systems are evolved HB
stars that spent their ZAHB phase on the blue side of the instability strip.
This may suggest that age may not be the primary explanation 
for the Oosterhoff types.  
\end{abstract}

\section{Introduction}
Within our Galaxy, most globular clusters that contain RR Lyrae 
variable stars can be divided into two distinct groups: in Oosterhoff I 
clusters the fundamental-mode RR Lyrae variables have an average 
period near 0.55 days, while in Oosterhoff II clusters the average 
fundamental-mode period is close to 0.65 days.  The two
types of cluster also differ in the relative numbers of first overtone 
pulsators (RR$1$-type variables) versus fundamental mode pulsators 
(RR0-type variables), and in their period-amplitude (PA) relations
\citep{leecarney99, pritzl00, catelan09}.  

The Oosterhoff dichotomy has been a source of great interest for 
decades in modern stellar astrophysics, especially after it was shown that our 
Galaxy is the only known example to exhibit such a dichotomy. The RR Lyraes 
in neighboring dwarf galaxies have different values of $\rm< P_{RR0} >$
than either of the OoI and OoII groups \citep[$e.g.$,][]{dallora03, catelan04, catelan09,
greco09}.  An understanding of the Oosterhoff effect 
is essential to any understanding of the formation and evolution of the globular 
cluster system and the formation history of the Galaxy \citep{eggen62, searle78} .  

The leading theoretical explanation for this phenomenon involves a
dichotomy in the ``transition period" between the RR0 and RR$1$ 
variables which reflects a difference in effective temperature at the 
transition point \citep{vanalbada73}.  This ``hysteresis effect" would 
cause a delay in the mode switching (from $e.g.$, RR0 to RR$1$) and 
would occur at different temperatures, depending on the direction of 
evolution.  According to this scenario, for the OoI, intermediate metallicity 
clusters, most of the horizontal-branch stars begin their lives in or near the 
instability strip.  The RR Lyraes become hotter and their periods become
shorter as they evolve and eventually the RR0
stars switch their mode to become RR$1$-type variables.  For the OoII, more
metal-poor clusters, the HB stars begin their lives on the hotter (blue)
side of the instability strip and subsequently evolve to lower temperatures 
and longer periods as
they traverse the instability strip at luminosities appreciably higher than
the zero-age horizontal branch.  Thus, a luminosity (mean density)
difference contributes to the effect in addition to the direction of evolution as
per van Albada and Baker.  As an example, \citet{bono97a} show evolutionary 
tracks of HB stars with the predicted instability strip and point to where the transition
is expected to take place according to the hysteresis hypothesis.

This explanation also has the advantage that it would help explain the 
larger fraction of RR$1$ variables in OoII  clusters.  The metallicity of the 
cluster is the ``first" parameter that establishes where on the horizontal 
branch the average star settles after core helium burning begins.  But the 
exact ZAHB location of each individual star is governed by the efficiency 
of the mass-loss along the RGB.  The more metal-poor clusters 
have bluer horizontal branches and would become OoII clusters and vice 
versa for the intermediate metallicity OoI clusters.  However, this explanation 
does not explain why no dichotomy is seen in the globular clusters belonging 
to dwarf spheroidal galaxies unless the range in metallicity of the old population 
in each galaxy is very small.  

Although it is common to categorize the GCs as OoI and OoII clusters, the 
Oosterhoff effect can manifest itself on a star-by-star basis, 
rather than only for the mean values of a given cluster 
(Sandage, Katem \& Sandage 1981).  The period-amplitude relations for RR 
Lyrae variables in OoI and OoII globular clusters differ substantially, and the 
position of an RR Lyrae in the PA plane is often used as a diagnostic to 
determine whether an RR Lyrae is OoI- or OoII-type.  For example,
it is not uncommon to find a few RR Lyrae variables in OoI globular clusters 
that occupy the OoII PA position and these are believed to be more evolved and 
intrinsically brighter than the majority of the OoI stars 
($e.g.$, Cacciari, Corwin \& Carney 2005, Sandage et~al. 1981).
If evolution is the explanation for the Oosterhoff groups, then this 
shift of the stars' positions in the PA plane by OoI and OoII RR Lyrae 
stars would be explained not by a simple difference in abundances, (see
Clement \& Shelton 1999), but by evolutionary effects.  However, it was 
shown that both the ``young" and ``old" halo GCs (using the
Mackey \& van den Bergh 2005 classification scheme) 
present the Oosterhoff dichotomy \citep{catelan09}, in contrast to the 
notion that OoII clusters as a group are more evolved and hence older 
than OoI clusters.  When using the (smaller-sample) 
Mar\'{i}n-Franch et~al. 2009 relative age classification scheme, this is not 
as clear.

Another explanation for the Oosterhoff groups is that helium abundance is affecting
the RR Lyrae variables in the OoI and OoII globular clusters.  It has recently
been shown by \citet{busso07} and \citet{caloi07} that a helium enriched population could 
explain the HBs in the peculiar, metal-rich globular clusters NGC 6388 and NGC 6441.  
These two metal-rich GCs ($\rm [Fe/H]$$\sim$$-$0.6 dex) have complicated the 
Oosterhoff groups since they have been found to contain RR Lyrae variables 
with unusually long periods.  In terms of the evolutionary
explanation for the Oosterhoff groups, these would be clearly classified as 
extreme Oosterhoff type II clusters.  However, it has been shown by 
\citet{pritzl00} that neither their HB morphology in the color-magnitude diagrams 
nor theoretical models by Sweigart \& Catelan (1998) indicate that the RR Lyraes 
are in a more advanced evolutionary stage.  Instead, the presence of multiple 
stellar populations with different initial He contents could account for the peculiar 
morphology of the HB in both clusters \citep{busso07}.  Further, it is seen that 
clusters with different metallicities exhibit similar HB morphologies, while 
clusters of the same metal abundance can have rather different HB morphologies.  
This is the ``second parameter'' puzzle, which also may be explained (at least 
in part) by differences in helium content  (see $e.g.$, Gratton et al. 2010).
 
As a by product of understanding the Oosterhoff effect, vital information on 
the second parameter phenomenon may be obtained.  We will here investigate 
the Oosterhoff dichotomy by comparing the 
evolutionary status of OoI and OoII RR Lyrae stars, where the evolutionary 
status is determined from the RR Lyrae period change rates.  In addition
we report on a pilot study to formulate Period-Color (PC) and 
Amplitude-Color (AC) relations as a function of pulsation phase and Oosterhoff type. 
These resulting relations could permit a more detailed comparison of RR Lyrae 
properties with Oosterhoff type. 

New period change rates are obtained for the RR Lyrae variables in IC$\,$4499.  
IC$\,$4499 is considered an OoI cluster, with a mean RR Lyrae period 
of $\rm <P_{RR0}>$=0.579 d and $\rm <P_{RR1}>$=0.347$\,$d, and 
a ratio of the number of RR$1$ stars to the total number of RR 
Lyrae of $\rm N_{RR1}/N_{RR}$=0.20.  The IC$\,$4499 period change rates are compared
to the period change rates in the OoI clusters M5, M3 and NGC$\,$7006,
the OoII clusters M15, M2, NGC$\,$5466, M22 and NGC$\,$5053, the field RR0 
Lyrae variables and the mixed
population cluster $\omega$~Cen.  Period change rates give insight on the
evolution of RR Lyrae variables.  The periods of the RR Lyrae variables 
should be increasing if the star evolves from blue to red 
in the HR diagram and decreasing if the star evolves from red to blue 
\citep{sandage57}.   Hence, if evolution is responsible for the difference 
between OoI and OoII-type RR Lyrae variables, those stars with OoI properties 
would evolve from red to blue and should have decreasing periods.  Similarly, 
those RR Lyrae stars with OoII properties should have periods that increase.  
However, RR Lyrae period changes are not fully explained by evolution \citep{rosino73}, 
and it is frequently seen that observed period changes are an order of
magnitude larger than would be expected from HB evolutionary models\citep{sweigart79}.

By comparing where the increasing and decreasing RR Lyrae period change rates fall
on the PA diagram, we can explore whether a relation exists between Oosterhoff type 
and period change rate (evolutionary state?).  Although mean RR Lyrae period 
change rates have been
calculated for different globular clusters of different Oosterhoff groups, the individual RR Lyrae
period change rates (and hence direction of evolution) as a function of position on the 
PA plane has not been explored.  

This paper first addresses the observational data for the RR Lyrae variables 
in IC$\,$4499 in \S2.  The determination of their period change rates follows 
in \S3 and the presentation of these period change rates is
given in \S4.  Next, in \S5, the dereddened $\rm (\vmr)$ colors and minimum light
temperatures of the IC$\,$4499 (OoI) RR0 variables are compared to the
RR0 in M15 (OoII).  In \S6, implications regarding the period change rates of 
RR0 Lyrae variables in the different Oosterhoff GCs are discussed.  
A summary is presented in \S7.

\section{Data and Photometry}

The present photometric data reduction for IC$\,$4499 comes from 
analysis of 1,338 individual CCD images in the direction
of IC$\,$4499 from 23 observing runs on five telescopes (Magellan/Baade
6.5m, CTIO Blanco 4m, ESO NTT 3.6m, CTIO 1.5m, and CTIO 0.9m).  
These data have been placed by one of the authors (PBS)
on the photometric system of Landolt (1992; see
Stetson 2000, 2005), and span a range of slightly more than 
22 years (1987 March 2 to 2009 April 3).  The
various cameras used had different physical sizes and different image scales,
therefore slightly different areas around IC$\,$4499 were acquired
and any given star may not appear in all images.  
The average number of independent observations for a star was
165 in $B$, 244 in $V$, 36 in $R$ and 90 in $I$.
More details on the data sets can be found in our paper discussing the IC$\,$4499 
color magnitude diagram (Walker et al, in preparation).

Our analysis of the RR Lyrae light curves also makes use of the original
photographic $B$ data from Clement, Dickens \& Bingham(1979).  The $V$ and $R$
photometry for M15 comes from \citet{silbermann95}.

\section{Phase Shift Diagrams and Period Changes}
Period changes of variable stars are usually determined with the use of 
an O-C diagram, where the $O$ stands for the observed time of a particular
phase of the light curve, while the $C$ stands for the calculated or predicted 
time of that same phase.  Generally the O-C curve is calculated at a star's
maximum phase, and this is the case here as well.  Nonlinear O-C
curves indicate a changing period, and a linear O-C relation with a non-zero
slope signifies an inaccuracy in the adopted period (see, for example,
Storm, Carney \& Beck 1991).  In particular, a second-order (parabolic) fit indicates a constant 
period change rate.  From the pulsation equation a change in the period 
of an RR Lyrae means the structure of the star (specifically, the mean
density) is changing.  Hence, the
period changes of RR Lyraes can be connected with different phases of 
horizontal-branch evolution.  Although steady evolutionary effects do not explain
the large period decreases or the random and/or abrupt period changes that 
are seen in some local and GC RR Lyraes, a period increases as a star evolves 
redward and decreases as it moves blueward.  Also, any HB star's luminosity evolves
monotonically brightward.  As an RR Lyrae's luminosity increases so does
its radius increase and its density decreases.  Therefore, a star that
evolves both blueward and brightward should display a relatively 
small period change; a star that evolves both redward and brightward should 
display a slightly larger rate of period increase.  See \S4.2 for more details.  

The period changes of the IC$\,$4499 stars are
investigated using the most accurate photometry (errors smaller than 0.05
mag).  The $V$ and $B$ filters were chosen from among the six available
($U$,$B$,$V$,$R$,$I$,$DDO51$) because the $V$ data are the most 
abundant, and the early photographic plates (needed to extend the baseline) 
were mostly taken through a $B$ filter.  \citet{lee99} show that for 
the globular cluster M2 the phase shifts between the $B$ and $V$ light curves 
are quite similar:  the mean phase shift between the $V$ and $B$ light curves is only 
0.0021 ($\pm$ 0.0039 rms error) for the 19 RR0 stars, and 0.0001 ($\pm$ 
0.0021 rms error) for the 12 RR$1$ stars.  \citet{jurcsik01} point out that 
this same difference is less than 1-2\%  of the period. 

\begin{figure}[htb]
\includegraphics[width=16cm]{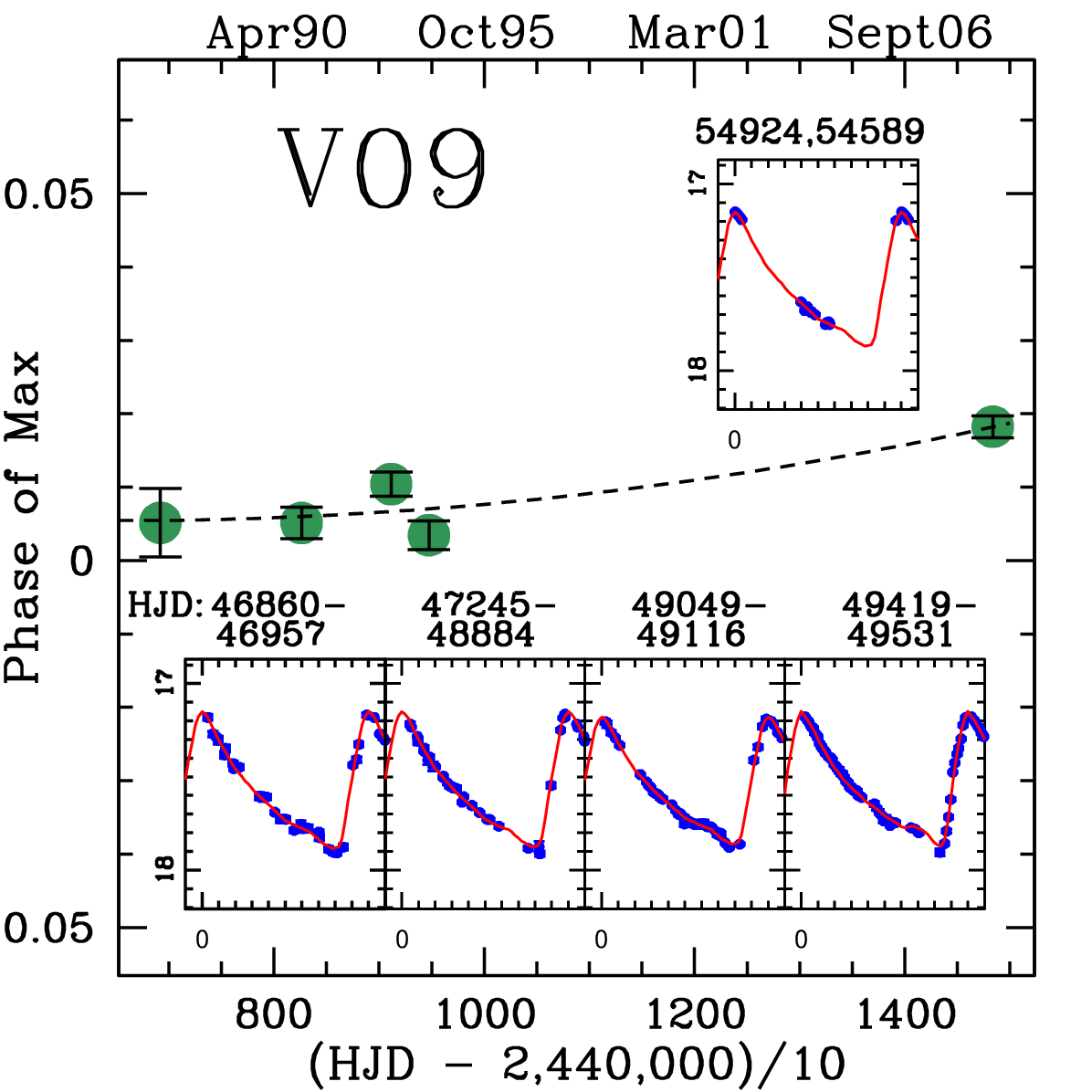}
\caption{An example of the construction on a typical O-C diagram using 
the $V$-band data of RR Lyrae V9 in IC$\,$4499.  The five different light
curves show the $V$-band data (represented by blue dots) in the 
five different time divisions.  The RR Lyrae template used to find 
$\phi$=0 is over-plotted in red for each light curve.
The HJD range is indicated above the lightcurves.  The light curve
with the most recent photometry, shown in the upper right hand corner, 
consists of data from only two nights.
The dashed line shows the weighted 2nd order polynomial fit.
\label{extemp}}
\end{figure}

To construct O-C diagrams, the $V$ photometric measurements were generally 
divided into five groups of time.  The $B$ photometric data were divided into 
either four or five groups of time, depending on the number of
\citet{clement79} observations.  In general, each group of data
blocked by time has $\sim$45
photometric values.  The data in each time group were folded by 
the period listed in Table~\ref{taboc} and shifted in phase until 
maxima aligned at $\phi$ = 0.  To help find where $\phi$=0.0, each light curve 
was fitted with a series of six light curve templates as described in \citet{layden98}.  
The resulting fit was visually examined to ensure that the fitted template 
accurately describes the light curve shapes, especially at maximum light.
An example of the quality and quantity of data in each time group, as well
as an example of the light curve templates' ability to align $\phi$=0,
are shown in Figure~\ref{extemp}.

Constructing an O-C diagram in this manner requires a relatively 
complete light curve for each time group.  Hence it can happen that
although there may be some data at a given time interval, if the
data are not evenly spaced about the light curve, no accurate maximum
light estimate can be made.  Although this limits the number of values plotted
in the O-C diagram, it also ensures that each point on the O-C curve
diagram is determined very accurately, regardless of whether the RR Lyrae
light curve shape is slowly changing.  For the RR Lyraes in IC$\,$4499
such a procedure works well, as the photometry can be easily divided
into groups of time with adequate coverage for each group 
(see Figure~\ref{extemp}).  The resulting period change rates were
compared to those obtained from one of the author's (JMN) SAS program 
(Nemec, Hazen-Liller \& Hesser 1985) and yielded consistent results.

There are two relatively large gaps in the data, which have the potential
to introduce cycle-count ambiguities.  Stars with shorter periods and larger
period change rates are most prone to this.  To check for this, 
each $V$ and $B$ light curve was divided at HJD = 2,449,000 and a direct 
period was calculated for these two groups using the RR Lyrae period finding 
approach developed by \citet{layden99}.  For the stars with the largest period 
changes and hence those with potential cycle-count ambiguities, 
the difference between the directly calculated periods
of the earlier and later HJD time groups 
would indicate whether the period change rates are in the right direction.
To calculate the O-C values
for these stars, cycle counts were assumed that produced values 
continuing the previous trend in the O-C diagram.  

\begin{figure}[htb]
\includegraphics[width=16cm]{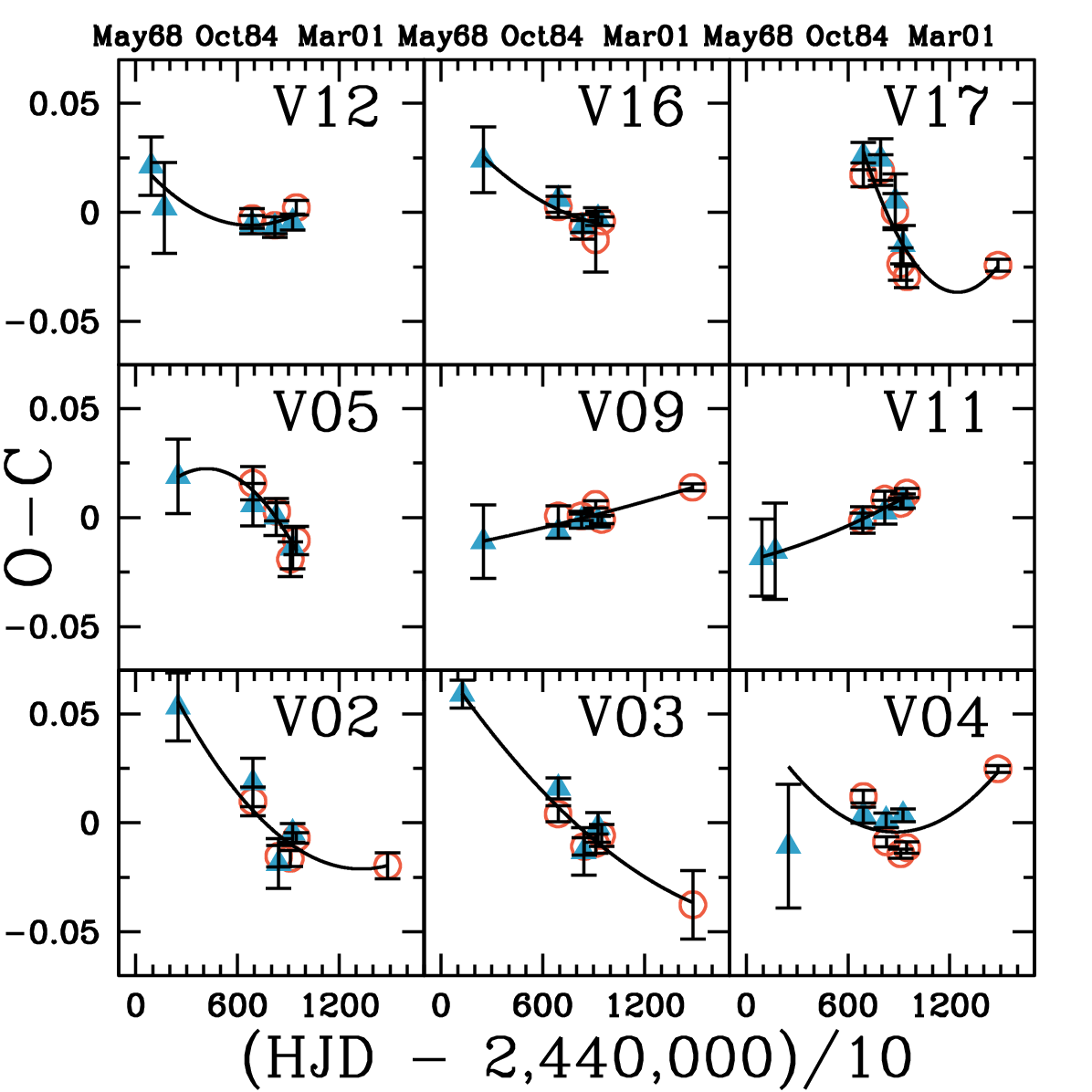}
\caption{The O-C diagrams for the RR Lyrae variables in IC$\,$4499.  The triangles
(blue) represent the phase shifts derived from the $B$-band data,
and the open circles (red) indicate the phase shifts derived from the
$V$-band data.
\label{OC1}}
\end{figure}

\begin{figure}[htb]
\includegraphics[width=16cm]{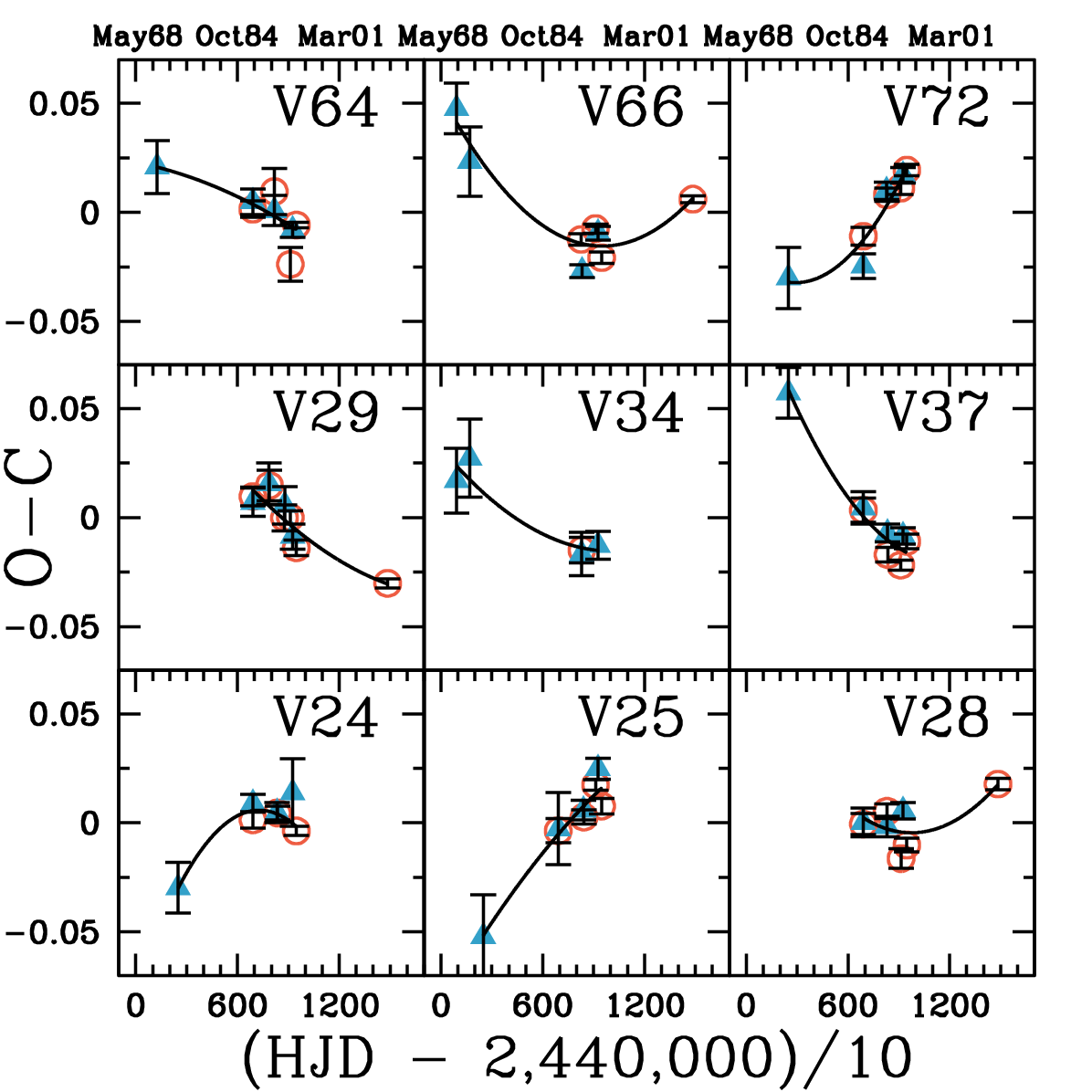}
\caption{The O-C diagrams for the RR Lyrae variables in IC$\,$4499.
Symbols are the same as in Figure~\ref{OC1}.
\label{OC2}}
\end{figure}

\begin{figure}[htb]
\includegraphics[width=16cm]{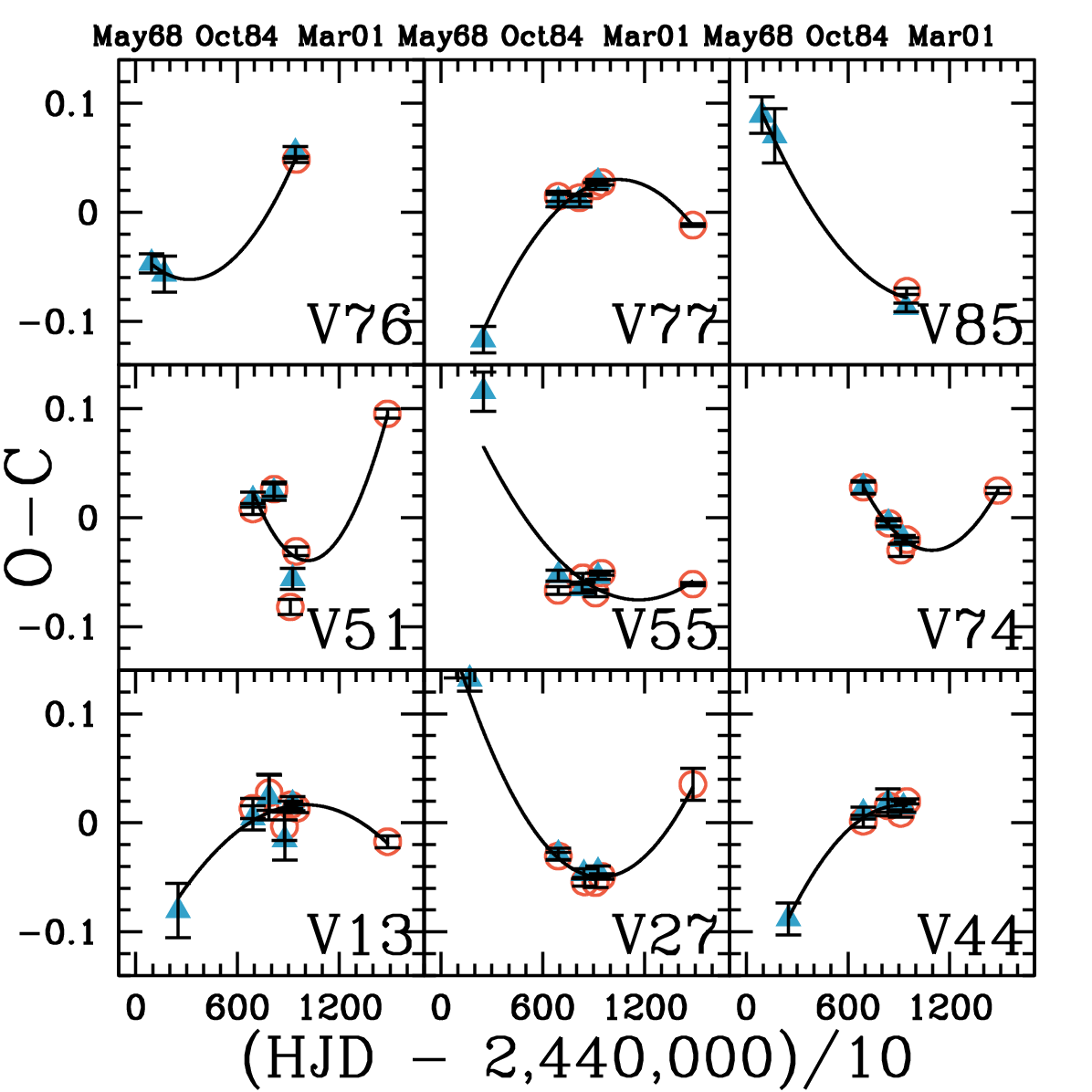}
\caption{The O-C diagrams for the RR Lyrae variables in IC$\,$4499.
\label{OC3}}
\end{figure}

\begin{figure}[htb]
\includegraphics[width=16cm]{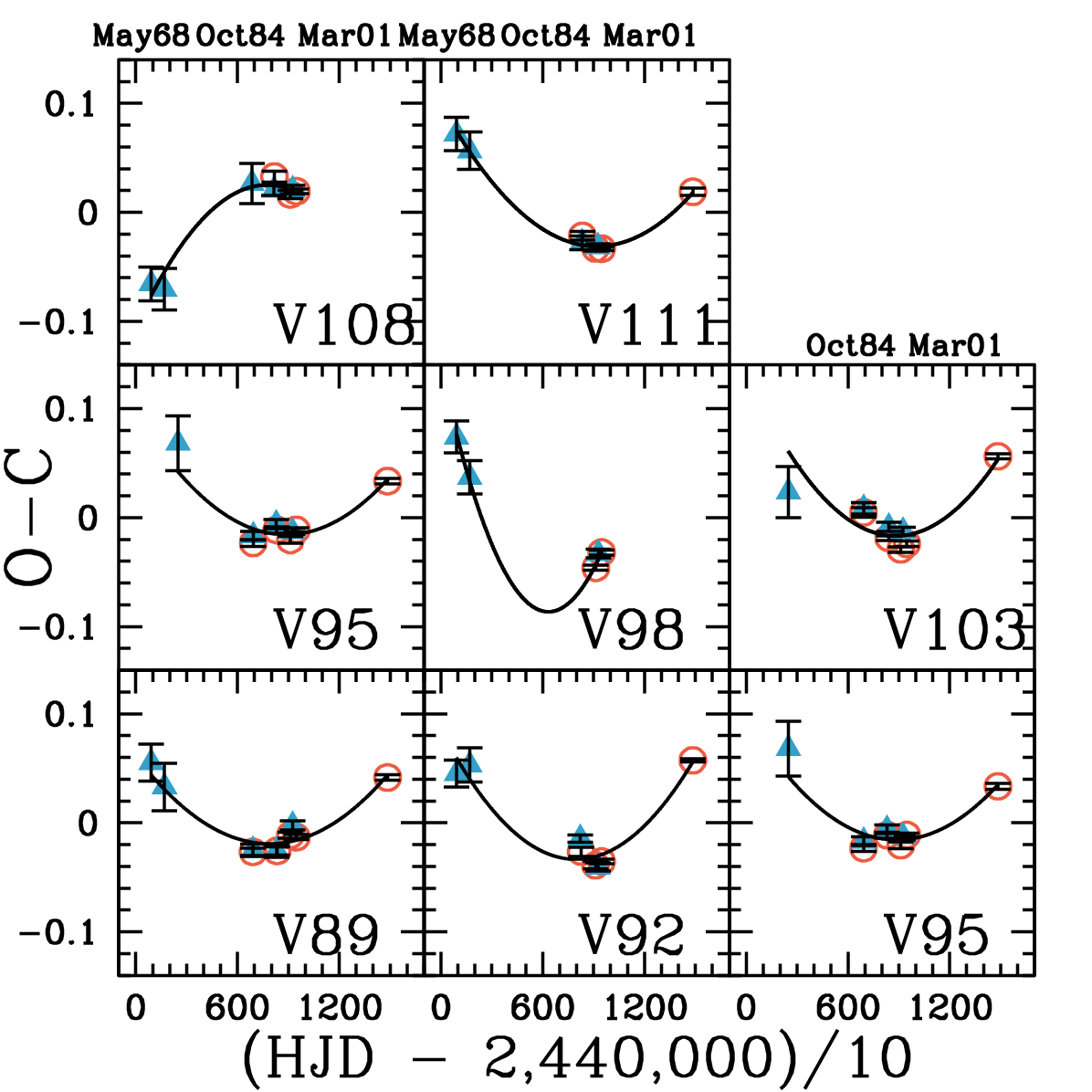}
\caption{The O-C diagrams for the RR Lyrae variables in IC$\,$4499.
\label{OC4}}
\end{figure}

\begin{figure}[htb]
\includegraphics[width=16cm]{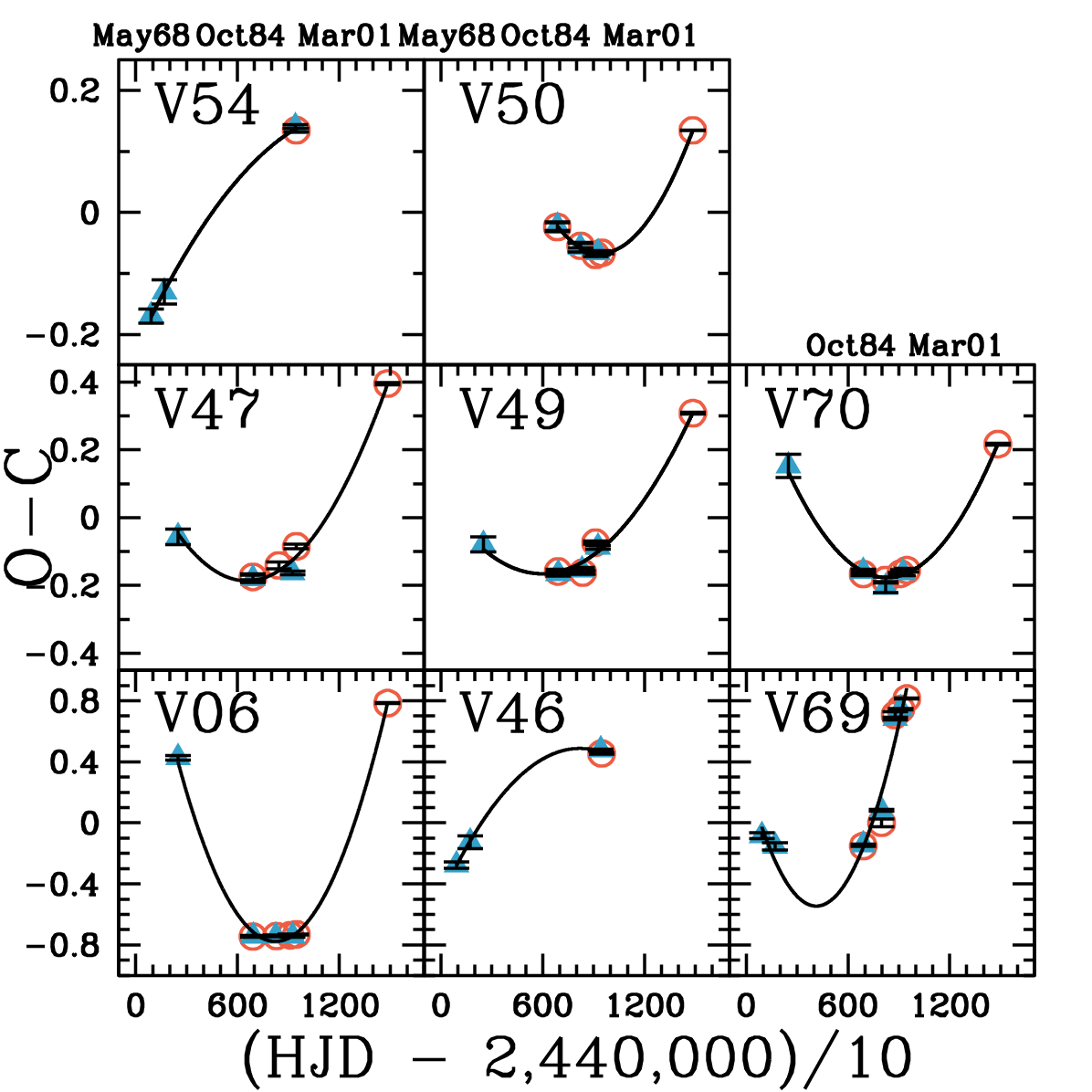}
\caption{The O-C diagrams for the RR Lyrae variables in IC$\,$4499
with relatively large period change rates.  
Note an increase in the scale of the vertical axis to encompass these
period change rates.
\label{OC5}}
\end{figure}

The shift required to align the light curve at $\phi$=0 (phase at maximum) is plotted 
against the mean HJD in Figure~\ref{OC1}-\ref{OC5}.   
A weighted second-order (parabolic) fit as well as the scatter about the 
fit are calculated for each star.  Following \citet{jurcsik01}, the stars with 
a period change rate, $<\Delta P / \Delta t>$, of

\begin{equation}
|<\Delta P / \Delta t> | \geq 2\sigma_{\Delta P / \Delta t}
\end{equation}
\noindent
are likely to have a monotonic period change.  Here $P$ is
the assumed period listed in Table~\ref{taboc}.  Table~\ref{taboc}
consists of (1)~the variable number, (2)~the adopted period, 
(3)~the values of $\beta$ in d Myr$^{-1}$, (4)~the
formal error in $\beta$ in d Myr$^{-1}$, (5)~the normalized 
period variations, $\alpha$ = $\beta$/$P_a$ in 10$^{-10}$d$^{-1}$, 
and (6)~the $\sigma_{\Delta P / \Delta t}$ $\times$ 2 in d Myr$^{-1}$
for all the variables studied.  

As the above procedure requires a thorough examination of the 
RR Lyrae light curves, it became evident that many stars exhibited 
Blazhko behavior (amplitude and/or phase modulation, see $e.g.$,
Jurcsik et~al. 2009, Kolenberg 2010).  Stars that 
exhibit the Blazhko behavior must be treated with special caution, 
as disentangling the Blazhko behavior from the period change 
behavior requires a lot of data and is not always straightforward.  
Although we attempted to solve for the period change rates for the 
Blazhko stars, the errors in $\beta$ are quite large.  More data would 
help beat down these errors.  We would be extremely grateful to any 
investigators willing to donate copies of proprietary CCD images or 
calibrated photometric measurements of IC$\,$4499 variable stars 
to enhance our data set for this cluster.  The 21 stars (31\% of the 
IC$\,$4499 RR0 population) that appear to exhibit Blazhko behavior 
are: V1, V15, V19, V20, V23, V30, V38, V43, V48, V52,
V53, V57, V58, V61, V74, V80, V82, V83, V84, V85, V88.  This is a 
larger percentage than seen in $\omega$ Cen \citep[20\% 
found by][]{jurcsik01}, and a smaller percentage 
than seen in the field RR0 stars \citep[47\% found by][]{jurcsik09}.

\section{Horizontal-Branch Evolution}
\subsection{Period Change Rates}
\begin{figure}[htb]
\includegraphics[width=16cm]{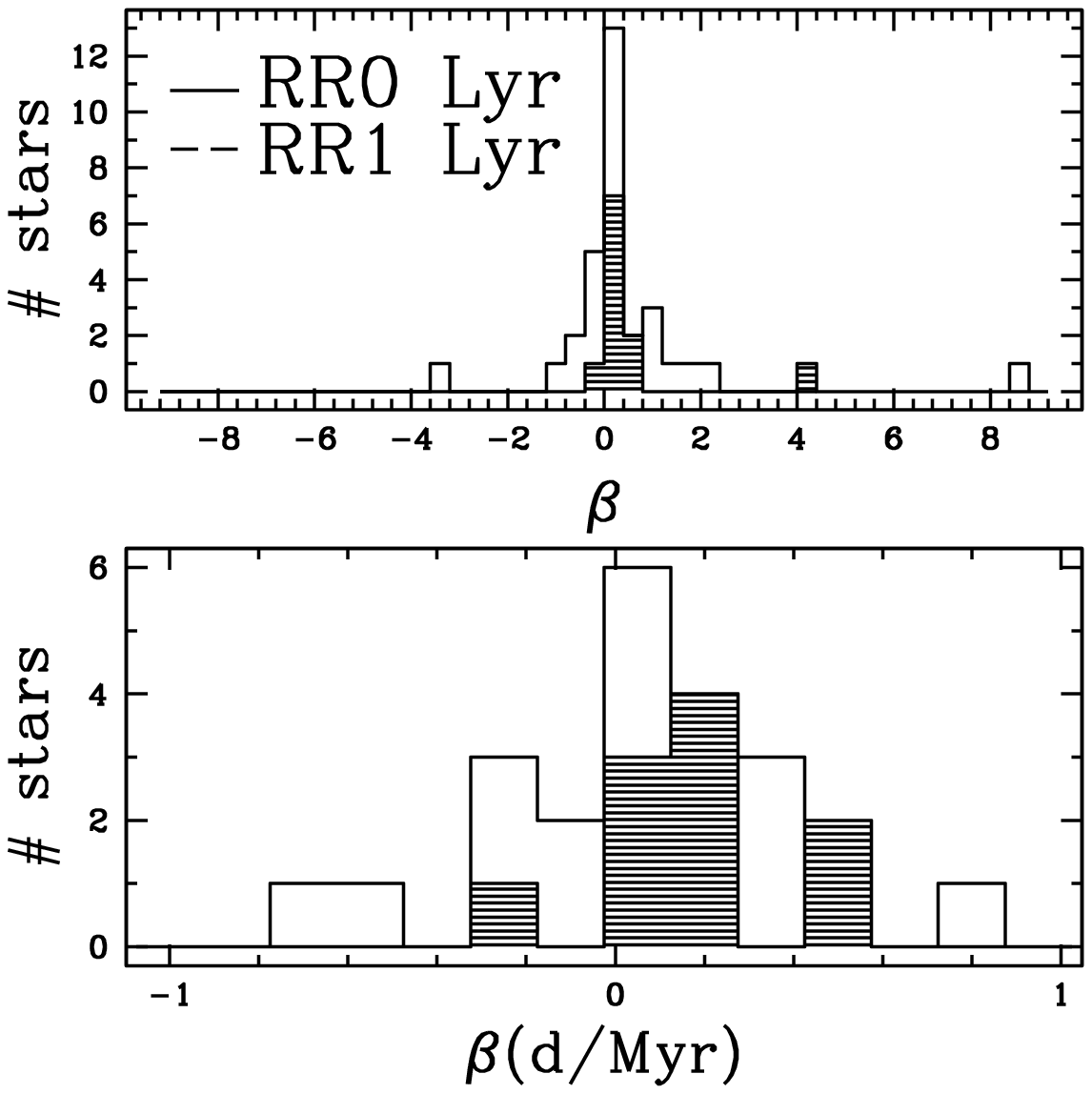}
\caption{The period change rate of 31 RR0 and 11 RR$1$-type variables in
IC$\,$4499.  The bin size is 0.4 d Myr$^{-1}$.  The $lower$ $panel$ presents the 
same data as the upper panel, but zooming in around the peak of the 
distribution.  Here the bin size is 0.15 d Myr$^{-1}$.
\label{HistBeta}}
\end{figure}

Figure~\ref{HistBeta} is a histogram of the 31 RR0 and the 11
RR$1$ variables that are fitted with parabolas that yield
period change rates that predict acceptable periods for phasing
our data.  It is striking that both RR0 Lyrae and 
RR$1$-type variables tend to have positive period change rates.  
The distribution of the period change rates for the IC$\,$4499 
RR Lyrae variables does not scatter evenly about zero, but there 
is a preference for $\beta$ values between 0.0 and 0.1 day 
Myr$^{-1}$.  There are 22 RR0 stars with positive period change 
rates and 9 RR0 with negative period change rates.  For the 
RR$1$-type variables, only one star with a negative period 
change rate is seen.  The stars not included in this figure are the 
possible RR2("RRe", second-overtone pulsation), 
RR01 ("RRd" double-mode pulsation), most of the probable Blazhko 
variables, stars that have ambiguous 
$O-C$ curves (V26,  V32, V38, V96, V106) and stars for which there were
not enough data to draw any firm conclusion about their long-term
period changes (V45, V46).  The general shape of the
histogram does not change upon removing the 12 RR Lyrae variables
(6 RR0 and 6 RR1 variables) with period change rates that
do not conform to Equation~1.  The ratio of stars with increasing
periods to total stars with calculated period change rates 
is approximately the same ($\sim$70\%).  

As seen in other studies of period change rates of RR Lyrae stars in GCs
\citep[$e.g.$,][]{reid96, silbermann95}, there is an excess of large period 
changes among the RR0 stars compared to the RR$1$ stars.  A plausible 
explanation for larger period changes among fundamental pulsators when 
compared with first overtones is explained in the topology of the 
instability strip.  The region where fundamentals are pulsationally 
unstable is a factor of 3-4 larger (in effective temperature) than for 
the first overtone. This would imply that you have a larger probability 
to detect period changes among fundamental RR Lyrae variables. 

%Unfortunately the errors in the period change rates are rather large; 
%for IC$\,$4499 a weighted average period change rate of 
%$<\beta>$ $=$ 0.004 $\pm$ 0.10 d Myr$^{-1}$ is found, where the error indicates the 
%standard error of the mean.  The errors used in the weighted mean
%$<\beta>$ are those listed in
%Table~1, column 4, which come from from the formal error in 
%$\beta$ as determined by the parabolic fit of the $O-C$ diagram. 
%Removing the three outliers with $|\beta|$ $>$ 
%2.5 d Myr$^{-1}$ gives a mean of $<\beta>$ $=$ 0.004 $\pm$ 0.07 d Myr$^{-1}$.
%The mean period change values for the RR0 and RR1 variables
%separately are $<\beta >_{RR0}$ $=$ 0.014 $\pm$ 0.12 d Myr$^{-1}$ and
%$<\beta >_{RR1}$ $=$ 0.004 $\pm$ 0.14 d Myr$^{-1}$, respectively.
%The mode of the distribution is $<\beta >_{mode}$ $=$ 0.13 $\pm$ 0.05 d 
%Myr$^{-1}$, where the error indicates the uncertainty in determining 
%the mode only.  

\subsection{Period Change Rates From Stellar Evolution}
The period changes discussed above can in principle be used to 
test models of HB evolution.  However, the fact remains that many 
RR Lyrae period change rates have been measured to be an order 
of magnitude larger than predicted from stellar evolution theory
in a variety of GCs from a variety of sources, and this is the case for
the RR Lyrae variables in IC$\,$4499 as well.   It has been shown that 
when finding period change rates 
for RR Lyrae stars, there can be ``noise" which can be an order of 
magnitude larger and is attributed not to a star's secular evolution 
but to, $e.g.$, mixing events in the stellar core \citep{sweigart79}, 
hydromagnetic events \citep{stothers80}, convection \citep{stothers10}, 
or passage through the RR Lyrae instability strip by stars which are 
pre-ZAHB pulsators \citep{silva08}.  However, \citet{bono97b} used
the metal-rich RR Lyrae to show that erratic period changes have timescales
significantly shorter than the evolutionary effects, so the
probability to detect objects during an erratic change is much smaller
then for the secular change.  %This is likely to be true for the
%metal-poor RR Lyrae as well.

\begin{figure}[htb]
\includegraphics[width=16cm]{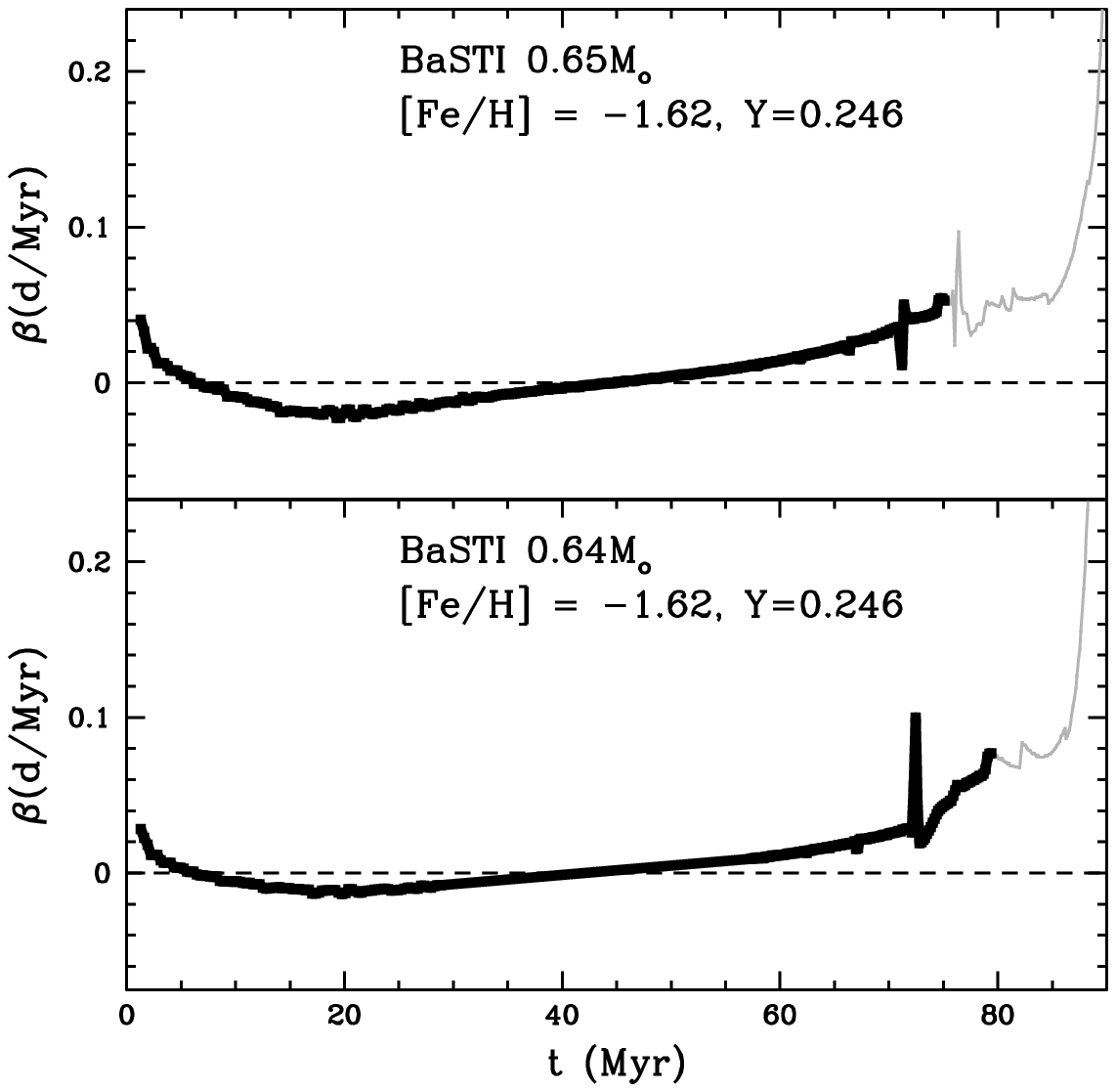}
\caption{The evolutionary period change rates ($\beta$ in d Myr$^{-1}$)
predicted for the horizontal branch stars belonging to IC$\,$4499 as a 
function of the time elapsed since the ZAHB.  The heavy lines indicate the
approximate range of the instability strip in IC$\,$4499, as observed by \citet{walker96}.
See text for details.
\label{PulEqnAge}}
\end{figure}

So why is there an abundance of RR Lyrae period change rates seen 
that are an order of magnitude larger than those predicted by HB 
models?  One possible explanation is that the predictions of the theoretical 
models are incorrect.  This was put forth by \citet{leborgne07}, who determined 
period change rates for local RR Lyrae variables from the 
GEOS RR Lyrae database--a database 
containing about 50$\,$000 maximum times from more than 3000 RR 
Lyrae variables.\footnote{The GEOS database is accessible at 
{\tt http://dbrr.ast.obs-mip.fr/}, hosted by the Laboratoire
dÕ Astrophysique de Toulouse-Tarbes, Observatoire Midi-Pyr\'{e}n\'{e}es, 
Toulouse, France.}  Their $O-C$ plots were produced from 
data spanning more than 50 years (and for 
several objects the data exceeded 100 years).  A large 
number of their RR0 Lyrae variables showed evidence of linear 
period variations, and the determined period change rates
were of the same order of magnitude, the median 
values being $\beta$ = +0.14 d Myr$^{-1}$ for the 27 stars with 
increasing periods and $\beta$ = $-$0.20 d Myr$^{-1}$ for the 21 
stars with decreasing periods.  They further note that although a 
$\beta$ value of 0.30 d Myr$^{-1}$ is exceptional
 and the lifetimes of these phases would be very short according to
 HB models (see also Fig~\ref{PulEqnAge}), in their sample, they had 18 
 stars with 0 $<$ $\beta$ $<$ 0.30 and 9 stars with $\beta$ $>$ 0.30.
This 2:1 ratio seems too small to be caused by the presence of exceptional 
cases. 
%The theoretical models should also match the observed 
%$\beta$ values in a more satisfactory way, as these seem to be 
%higher than expected, both for redward and blueward evolutions.

Another explanation for the abundance of period change rates 
observed that are an order of magnitude larger than predicted 
by theoretical models, was put forth by \citet{lee91}.
They demonstrate that horizontal branch simulations can reproduce 
the observed distributions of $\beta$ in GCs if random observational 
errors are superposed on the evolutionary period changes.  
Hence the observed RR Lyrae period change rates can be attributed to 
evolutionary effects if the random observational error is of order 
$\pm$0.07 d Myr$^{-1}$ in $\beta$, a value consistent with those 
suggested by the observers.  
%We repeat a similar analysis for
%IC$\,$4499.  
Although the more recent 
HB models show basically the same behavior as those used
by \citet{lee91}, recent observational investigations present
period change rates with smaller observational errors
\citep{jurcsik01, nemec04, leborgne07}.
In general, this is the case for the observed period change rates 
presented here.  It is difficult to understand where and how such
$\pm$0.07 d Myr$^{-1}$ errors in $\beta$ would arise from these
newer observations.

\begin{figure}[htb]
\includegraphics[width=16cm]{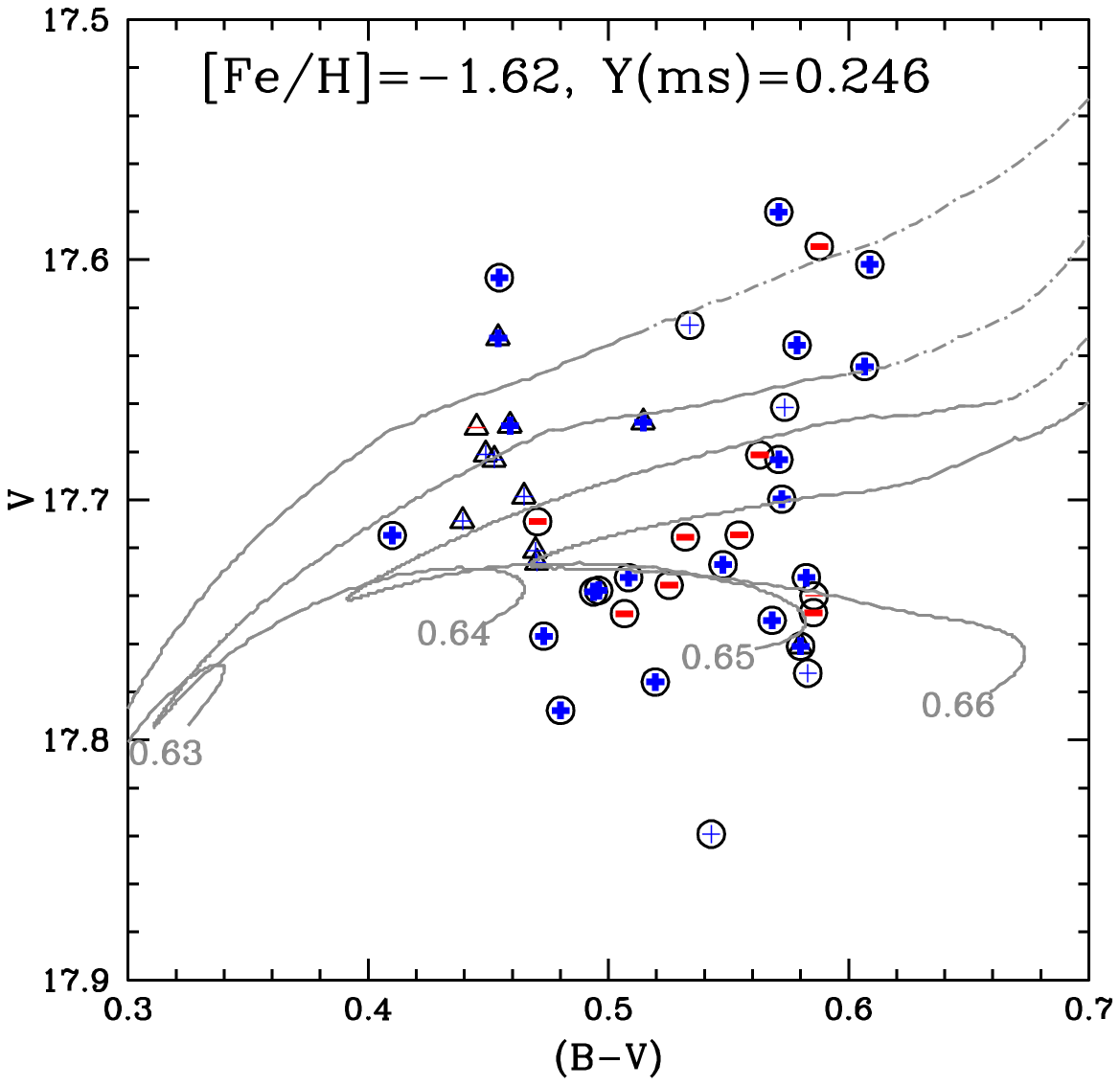}
\caption{$V$,$B-V$ CMD of the RR Lyrae for which period change rates
were determined.  The RR$1$ variables
are indicated as triangles and the RR0 variables are indicated
by circles.  The stars with positive period change rates are indicated by
plus signs, the stars with negative period
change rates are indicated by minus signs.  
Stars with period change rates that do not
conform to Equation~1 are indicated by light plus and
minus signs.  BaSTI HB evolutionary model models are plotted, with
$\rm [Fe/H]$=$-$1.62 dex, alpha-enhanced.  See text for details.
\label{cmd4499_2}}
\end{figure}

In Figure~\ref{cmd4499_2}, the $V$,$B-V$ CMD for all the stars
for which period change rates are determined is shown.  The evolutionary 
tracks for 0.63$M_{\sun}$ to 0.66$M_{\sun}$ from the BaSTI archive 
\citep{pietrinferni04, pietrinferni06} are also plotted,
where the tracks were calculated with $\rm [Fe/H]$ = $-$1.62 dex and
$\rm [\alpha/Fe]$ $=$ 0.4 dex \citep[$e.g.$,][]{walker96}, and a mass loss rate 
$\eta$$=$0.4.\footnote{ The BaSTI archive is 
available on-line at {\tt http://albione.oa-teramo.inaf.it/}}
The central He exhaustion for the adopted chemical composition
is represented by the solid portion of the individual HB models.
We adopt $E(B-V)$=0.23, $M_{V}$=0.52 and
$<V_{\rm ZAHB}>$=17.72 \citep{walker96} to align the models from the
BaSTI archive to
the observed CMD.  The BaSTI tracks indicate that most of the IC$\,$4499 RR Lyrae 
stars have masses of 0.64$M_{\sun}$ and 0.65$M_{\sun}$.  The general shape of the
evolutionary tracks changes little assuming a slightly more metal-poor HB
model of $\rm [Fe/H]$ = $-$1.82 dex, although the masses that
fit the observed colors increases by $\sim$0.03$M_{\sun}$.  For both of these 
metallicities, the predicted masses are
in agreement with the pulsational masses of 0.65-0.70$M_{\sun}$ 
found from the RR01 (RRd) variables \citep{walker98}.  However, the
$\rm [Fe/H]$ = $-$1.62 dex HB models yield RR0 Lyrae 
periods which are in much better agreement to those found observationally
for the IC$\,$4499 RR Lyrae variables.  For the $\rm [Fe/H]$ = $-$1.82 dex tracks
and the mass range that fit the observations, the RR Lyrae 
periods are longer by $\sim$0.1 days.

In Figure~\ref{PulEqnAge} the evolutionary rates of
period change in d Myr$^{-1}$ are plotted as a function of the time elapsed since the
ZAHB.  To estimate $\beta$, the fundamental pulsation
equation of \citet{bono97a}: 

\begin{equation}
{\rm log}~P_F = 11.627 + 0.823~{\rm log}~L - 0.582~{\rm log}~M - 3.506~{\rm log}~T_{eff}
\end{equation}
has been used.  The heavy lines represent the
observed range of the IC$\,$4499 instability strip (0.115 $>$ $(B-V)_0$ $>$ 0.365; see
Walker \& Nemec 1996) for each track.  The evolutionary tracks run first
redward (see Fig~\ref{cmd4499_2}) giving positive period change rates
of ~0.03 d Myr$^{-1}$.   Soon, blueward evolution through the IS occurs, relatively
slowly, giving small negative rates of period change.  Faster positive
period changes occur as the redward evolution proceeds, yielding
$\beta$ values of up to $\sim$0.1 d Myr$^{-1}$.  These values are obviously
much smaller than the observed values for IC$\,$4499.

\begin{figure}[htb]
\includegraphics[width=16cm]{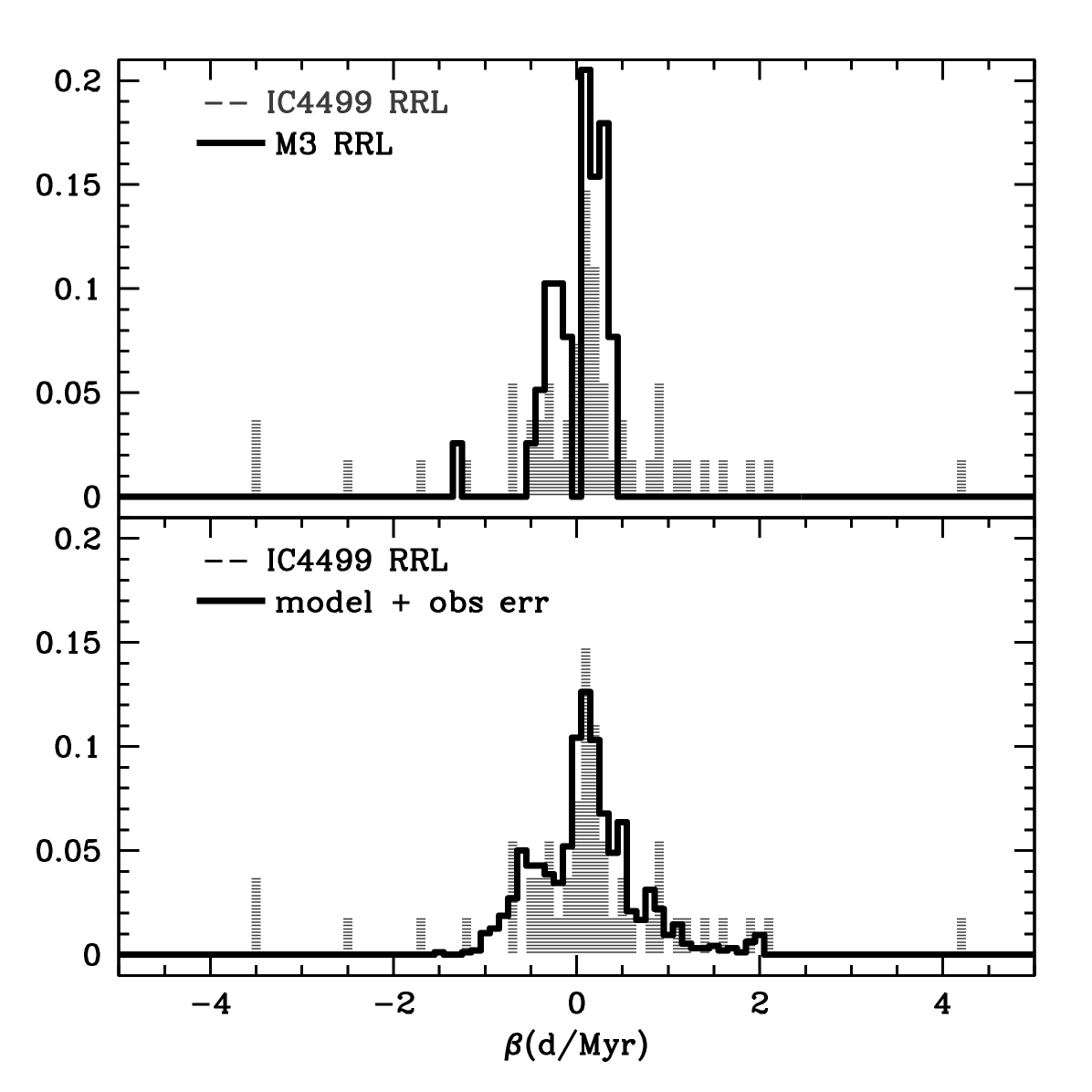}
\caption{  {\it Top:}  Comparison of the observed period
change rates of IC$\,$4499 presented here and M3 from \citet{corwin01}.
{\it Bottom:}  Comparison of the observed period change rates with those
predicted from evolutionary models, where the evolutionary models include
an observational error of $\pm$ 0.07 d Myr$^{-1}$.  The histograms
have been normalized.  
\label{dPdTpred}}
\end{figure}

In the bottom panel of Figure~\ref{dPdTpred} the histogram
of the evolutionary period change rates are presented from the
$\rm [Fe/H]$ = $-$1.62 dex HB tracks discussed above, assuming 
15\% of the RR Lyrae variables have 0.63$M_{\sun}$ and 
0.65$M_{\sun}$  and 35\% of the variables have 0.64$M_{\sun}$ 
and 0.65$M_{\sun}$ (see Fig~\ref{cmd4499_2}).  
To simulate observational errors, the evolutionary period change 
rates are enhanced by an order of magnitude 
($\sim$$\pm$0.07 d Myr$^{-1}$).  Because the baseline of the 
IC$\,$4499 observations is quite long, it is rather 
straightforward to observe if the period of
an RR Lyrae star either increases or decreases over time.
For this cluster, the addition of random observational errors, as introduced 
by \citet{lee91}, would not make sense.  Instead we introduce
a systematic error, as could be imagined for example, if
the errors on the different O-C points were underestimated and hence
the weighted parabolic fit would yield greater coefficients than
realistic.  As with \citet{lee91}, we find that the observed period change 
rate histogram is very similar to that predicted by HB models.  
Repeating this procedure using the $\rm [Fe/H]$ = $-$1.82 dex HB 
tracks with masses appropriate for the observed IC$\,$4499
observations, a similar shaped histogram is obtained but with a
slightly wider base.  This is not surprising given the similarity
in the shape of the evolutionary tracks at these two metallicities and
that the evolutionary period change rates of more metal poor
RR Lyrae variables tend to be larger.  

In the top panel of Figure~\ref{dPdTpred} the period change rates of the
IC$\,$4499 RR Lyrae are compared to those in M3, because M3 has a similar 
HB parameter and metallicity to IC$\,$4499.  The histograms look very
similar.  In particular, the period change rates are of the same order of magnitude.
As pointed out by \citet{leborgne07}
since almost all of RR Lyrae period change rates have the same
order of magnitude they should have a common origin.
It would be very unlikely that all the 42 RR Lyrae variable in IC$\,$4499 and
37 RR Lyrae variables in M3 for which
period variations are determined, are undergoing period 
variations caused by various instabilities, tidal effects, etc.
However, we cannot rule out the possibility that some individual 
stars in the 42 IC$\,$4499 and 37 M3 RR Lyrae with period change rates are merely
in a particular transition phase.  It is much more likely that the 
period changes for the RR Lyrae in IC$\,$4499 and M3 are caused by a 
more general phenomenon, with the immediate answer from \citet{leborgne07}
being long-term stellar evolution.  
Our period change rate values are similar to those reported by (the 
more recent papers only) \citet{jurcsik01} in the case of $\omega$~Cen 
RR Lyrae stars, \citet{leborgne07} in the case of the field RR Lyrae, 
\citet{corwin01} in the case of M3, and \citet{nemec04} in case of 
NGC 5053, and discussed as evolutionary changes. 

In conclusion, we can not prove that the period changes are due to 
evolution since the comparison with the models are poor.  The HB 
models predict period change rates that are an order of magnitude 
smaller than what is observed.  Given the distribution of the errors 
in the analysis, the
analysis by \citet{lee91}, which suggests an observational error
of $\pm$0.07 d Myr$^{-1}$ in $\beta$, does not seem valid, either.
Recent observed period changes have errors smaller
than what is required to match the observations and models in a
satisfactory manner.  The theoretical models should be fine 
tuned to match the observed rates in a better way, now that there 
are extensive databases available that allow 
period change determinations.

\subsection{The Period-Amplitude plane}
The Period-$V$-Amplitude ($\rm PA_V$) plane is often used to distinguish 
between OoI and OoII globular clusters.  Here 
we explore how the period change rates of the RR Lyrae variables in OoI 
and OoII globular clusters compare in the $\rm PA_V$ plane on 
a star-by-star basis.
% given that if RR Lyraes in OoI globular clusters were 
% evolving from red to blue, their periods would be predominantly decreasing,
% and vice versa for OoII globular clusters.  

%Observational studies of RR Lyraes in GCs have tended to show 
%$< \beta >$=0 for OoI globular clusters and $< \beta >$ $> $0 for OoII 
%clusters \citep{silbermann95, catelan09}.  It is worth noting that the
%uncertainties in many reported period change rates are large
%(often larger than the $\beta$ value itself) and the average period 
%change rates are generally not the weighted mean, which would 
%help compensate for those stars which, within the uncertainty limits, 
%could have either positive or negative period change rates.  

\begin{figure}[htb]
\includegraphics[width=16cm]{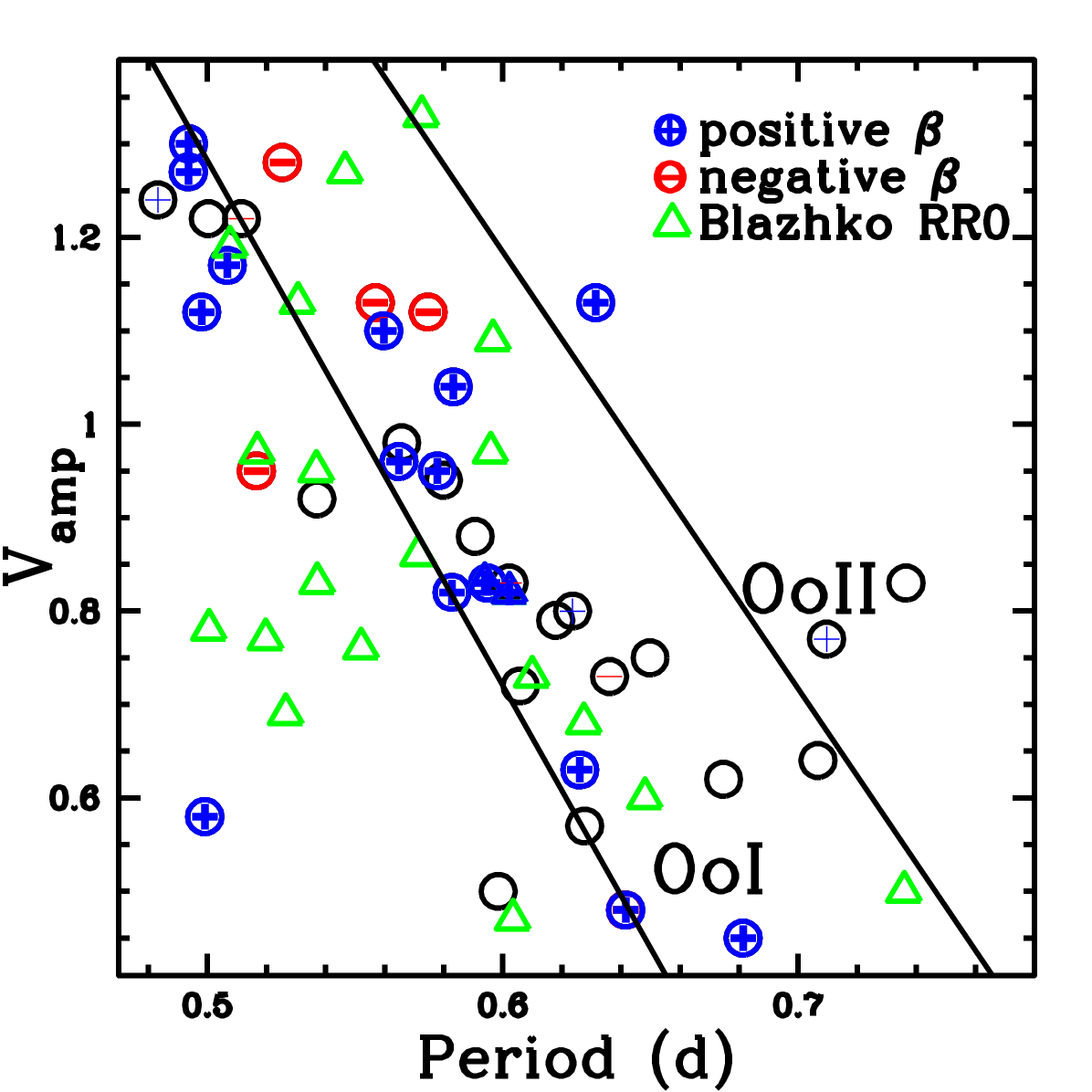} 
\caption{Period-amplitude relation for the RR0 variables in IC$\,$4499.
The stars with positive period change rates are indicated by
plus signs, the stars with negative period
change rates are indicated by minus signs, and the RR0 Lyrae
variables for which no period change rates were determined
are circles.  Stars with period change rates that do not
conform to Equation~1 are indicated by light plus and
minus signs.  The 
triangles denote stars that show evidence for the
Blazhko effect.  The least-squares fit to RR0 stars in the OoI prototype 
GC M3 and the least-squares fit to RR0 stars in the OoII GC M9 
from \citet{clement99} are indicated by solid lines.
\label{PAic4499}}
\end{figure}

Figure~\ref{PAic4499} shows the $\rm PA_V$ plane
for the IC$\,$4499 RR0 variables.  As IC$\,$4499 is an OoI cluster, more
RR Lyrae variables fall on the OoI line.  If RR Lyraes in OoI globular clusters were 
evolving from red to blue, their periods would be predominantly decreasing,
and vice versa for OoII GCs.  However, most of these 
OoI RR Lyraes have positive period change rates.  Here the OoI and OoII 
lines are defined by \citet{clement99}; the OoI line is the least-squares fit to 
RR0 stars in the OoI prototype GC M3, and the OoII line is the least-squares 
fit to RR0 stars in the OoII GC M9.

Blazhko variables tend to introduce scatter in the $\rm PA_V$ plane
due to their non-repeating light curves, and hence
their amplitudes vary over timescales longer than the basic pulsation period.
The Blazhko variables are denoted by triangles, and the RR0 variables
for which period change rates could not be determined are also
indicated.  By and large, these stars fall in the same region of the 
$\rm PA_V$ plane as the stars with well determined period change rates.
Hence a bias in which stars were sampled is minimal.

\begin{figure}[htb]
\includegraphics[width=16cm]{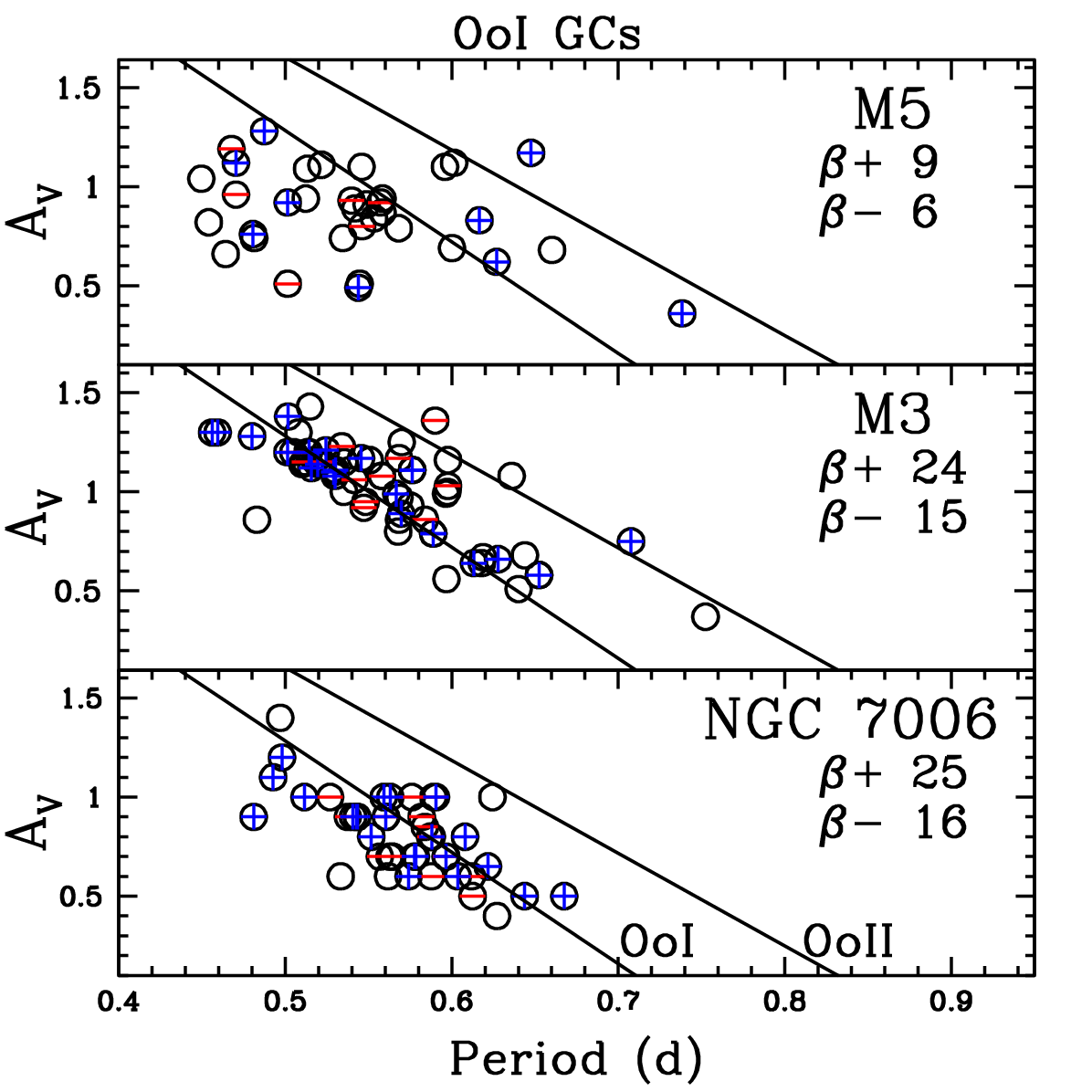}
\caption{Period-amplitude relation for the RR0 variables
in the OoI globular clusters M5, M3, and NGC 7006.
The stars with positive period change rates are indicated by
plus signs, and the RR0 variables with negative period
change rates are indicated by minus signs.  The stars
for which the period changes are unknown (or very uncertain)
are indicated by open circles. The OoI and OoII lines are taken 
from \citet{clement99}, determined from RR0 stars in 
the OoI GC M3 and the OoII GC M9.
\label{PAooI}}
\end{figure}

The $\rm PA_V$ diagram of the RR0 variables in the Oosterhoff I GCs M5, M3
and NGC$\,$7006 is shown in Figure~\ref{PAooI}.  The period change rates 
come from \citet{reid96}, \citet{corwin01} and Wehlau, Slawson \& Nemec (1999).
Again, there is a preponderant number of stars evolving from blue to red; on average,
there are 1.5 times as many RR0 variables with positive period change rates
compared to negative period change rates.  There does not appear
to be a trend of where the RR0 variables with either positive or negative 
period change rates lie.

Stars evolving from red to blue (negative $\beta$ values) along the horizontal
branch may have a period change rate approximately four times slower than
those moving in the reverse direction \citep{lee90}.  Hence, it could be that many stars with
no measured period change rates have negative $\beta$ values that are too small to
be measured.  Although this possibility can not be ruled out, even if all the stars 
with no determined period change rates had a negative period change rate, 
it is still striking how many RR Lyrae variables with positive period change rates exist
in the OoI clusters.  HB stellar models with metallicity suitable for OoI GCs show the 
blue excursion responsible for the negative period change only for a short interval 
of their HB lifetime.  As seen in \S4.2 for the IC$\,$4499 variables, the blue
excursion is $\sim$5 Myr (6\%) of their HB lifetime.

\begin{figure}[htb]
\includegraphics[width=16cm]{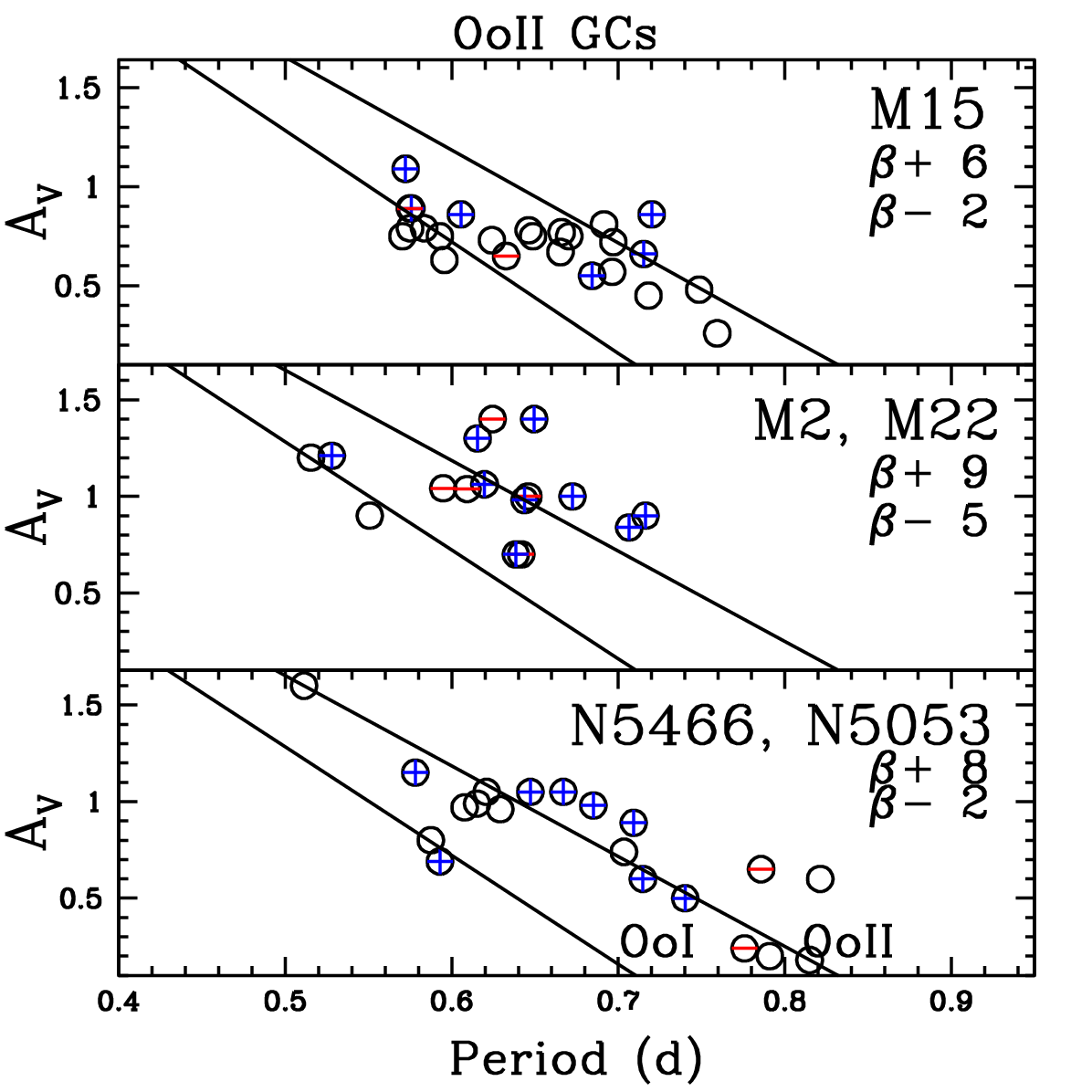} 
\caption{Period-amplitude relation for the RR0 variables in the OoII
globular clusters M15, M2, and NGC 5053.
The symbols are the same as in Figure~\ref{PAooI}.
\label{PAooII}}
\end{figure}

In Figure~\ref{PAooII}, the RR0 variables in the OoII GCs M2, NGC$\,$5466, 
M15, M22 and  NGC$\,$5053 are shown on a $\rm PA_V$ diagram.  The 
period change rates come from \citet{lee99}, Corwin, Carney \& Nifong (1999), 
\citet{silbermann95}, \citet{wehlau78} and \citet{nemec04}, respectively.
Although there are fewer RR0 variables with period change rate determinations,
it is clear that there are proportionately more stars with positive period
change rates than negative ones.  As with the OoI globular clusters, most RR0 
variables appear to have positive period change rates.  If the period change rates
can be attributed to evolution, this would mean both OoI and OoII RR Lyrae
variables would be evolving from blue to red.

Recently, high quality period change rates based on a plethora 
of data and long ($\sim$100 year) baselines have been determined
for RR Lyraes in the globular cluster $\omega$~Cen and
in the field by \citet{jurcsik01} and \citet{leborgne07}, respectively.  
Neither the Milky Way field, nor $\omega$ Cen is formally
seen as an OoI- or OoII-type object.  Rather, it has been suggested that $\omega$~Cen 
may be the remnant of a dwarf galaxy that was accreted by the 
main body of the Milky Way long ago \citep[$e.g.$,][]{norris96, meza05, dacosta08}.
The field RR Lyraes in the Galactic halo seem to display the 
separation into the two different Oosterhoff 
groups \citep{suntzeff91, bono97c, miceli08, delee08, szczygiel09}.  

\begin{figure}[htb]
\includegraphics[width=16cm]{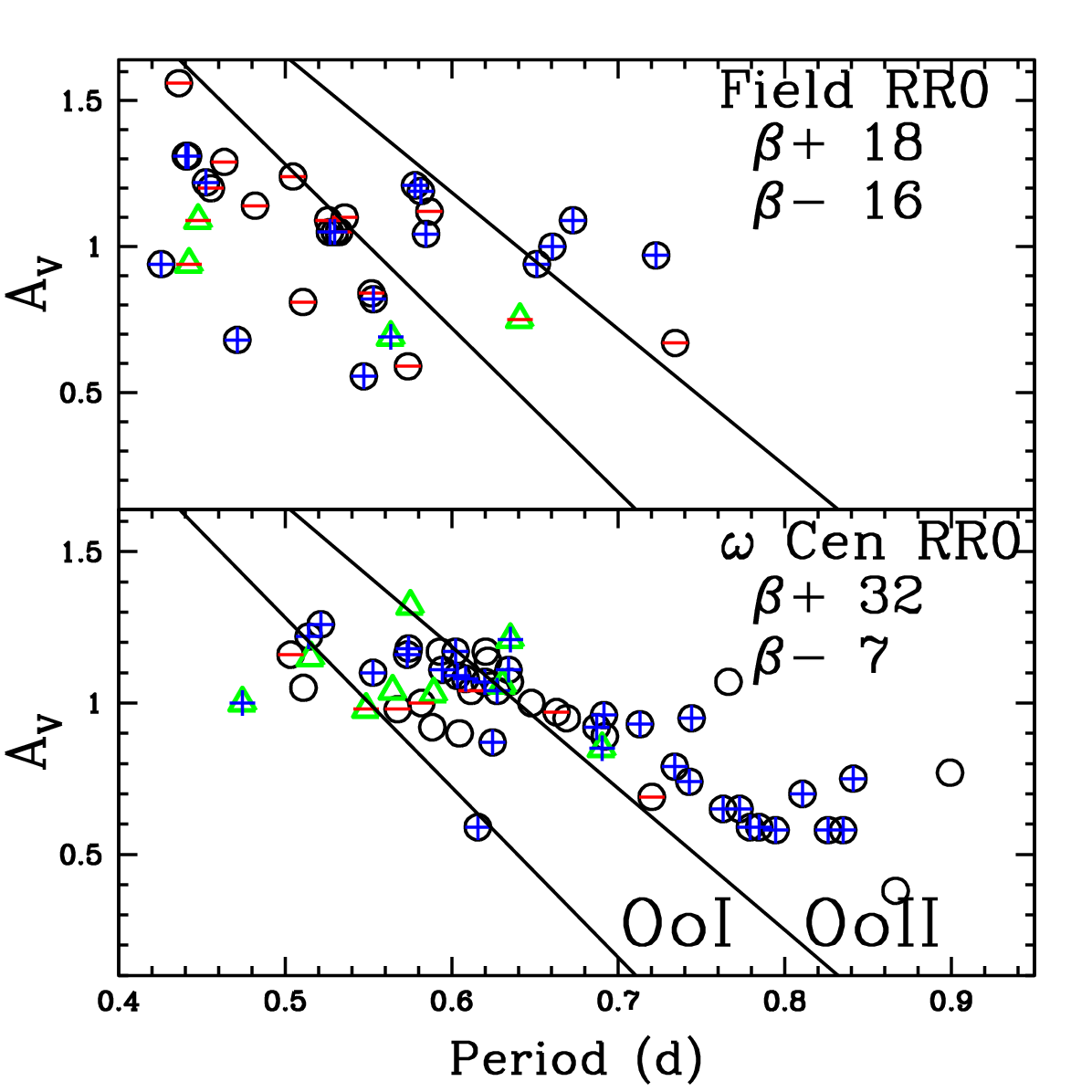} 
\caption{Period-amplitude relation for the RR0 variables in the mixed
population $\omega$ Cen and the field RR0 variables. 
The symbols are the same as in Figure~\ref{PAooI}.
\label{PAooQ}}
\end{figure}

The $\rm PA_V$ stars from these two studies are shown in Figure~\ref{PAooQ}.
Among the field RR Lyraes, blueward evolution (decreasing periods) is 
as common as redward (increasing periods).  For $\omega$~Cen,
period increases dominate the RR Lyrae population.  The RR0 variables
with negative period change rates again do not fall preferentially on either the OoI or OoII line,
but are scattered randomly in the $\rm PA_V$ plane.  Again, there does not
seem to be a relation between position on the $\rm PA_V$ plane (Oosterhoff type)
and period change rate (evolution?).

To quantify any differences between period change rates and 
Oosterhoff types, the mean, median and mode period changes 
for stars in two different parts of the $\rm PA_V$ diagrams are derived.
We define the OoI section, $\rm OoI-{PA_V}$, as:
\begin{equation}
A_V < (-0.73/0.13)~P + 4.3 
\end{equation}
which consists of variables closer to the OoI line in the $\rm PA_V$ diagram.
The OoII section, $\rm OoII-{PA_V}$:
\begin{equation}
A_V > (-0.73/0.13)~P + 4.3 
\end{equation}
consists of the variables closer to the OoII line.

Table~2 displays the results for the clusters studied above and
the Oosterhoff sections with $\sim$10 RR0 Lyrae variables. 
In Table~2, we list
%(the OoI GCs include
%M5, M3 and NGC 7006, and the OoII GCs include M15, M2, M22, NGC 5466
%and NGC 5053)
(1)~the particular population, GC or group of GCs, (2)~the Oosterhoff
section as defined by Equations~3 and ~4, (3)~the mean period
change rate and the error in the mean, in d Myr$^{-1}$,
(4)~the median value of $\beta$ in d Myr$^{-1}$, (5)~the
mode of $\beta$ in d Myr$^{-1}$, where the bin size used
in determining the mode is typically 0.05 d Myr$^{-1}$, and 
(6)~the number stars in each Oosterhoff section.
In general, the RR0 Lyrae variables in all the populations, GCs 
and Oosterhoff sections tend to have increasing period change rates.  
Moreover, the size of the period change rates are very similar for 
all the different Oosterhoff sections.  There is no indication of
correlation with period change rate and position in the $\rm PA_V$ plane.

\section{De-reddened Colors}
A search for differences in the intrinsic temperatures between OoI and OoII RR Lyraes
is also carried out using their $\rm(\vmr)$ colors at minimum light, $\rm (\vmr)_{min,0}$.   
Not only is $\rm(\vmr)$ known to be a good temperature indicator for variable stars
\citep{barnes76, barnes77, manduca81}, but \citet{kunder10} also showed 
that the $\rm (\vmr)_{min,0}$ for RR0 variables is constant 
to 0.02 mag and is largely independent of period, amplitude and metallicity.  
Therefore, $\rm (\vmr)_{min,0}$ can be used as in indicator
to probe for differences in the temperatures of OoI and OoII variables.
For example, it is well-known that OoI clusters tend to be more metal-rich and 
host fainter RR Lyraes than OoII clusters \citep{caputo00}.  
As the metallicity has an effect on the absolute magnitude of an RR Lyrae, 
it has been difficult to disentangle whether the metallicity difference alone is affecting 
the brightness differences, or whether there are differences in the intrinsic 
magnitudes of RR Lyrae variables in OoI and OoII
globular clusters caused by something other than just metallicity (like evolution
or helium).  Unfortunately, there is little $V$- and $R$-band data for RR Lyrae
variables in globular clusters to find their minimum light $\rm (\vmr)$ color.  Using new
$V$ and $R$ data of RR Lyrae variables in IC$\,$4499, we investigate
$\rm (\vmr)_{min,0}$ for the OoI cluster IC$\,$4499 compared to the OoII cluster M15.  

%Using the pulsation equation of van Albada \& Baker (1973), it was shown that the 
%longer periods of the OoII RR0 variables are caused by their higher 
%luminosities \citep{sandageea81, sandage81, sandage82}.
%Since RR Lyrae stars are used as distance indicators, this higher luminosity 
%should be accounted for when determining their absolute magnitudes.
%Indeed, As OoII clusters are generally more metal-poor than OoI clusters, the assumed
%notion that an RR Lyrae $\rm M^{\mathrm{RR}\,}_{V}$ depends primarily on \feh
%accounted for this enhanced luminosity of OoII variables.  
%Hence, it was still assumed that the dependence of the ZAHB luminosity 
%on \feh plays the largest role in an RR Lyrae absolute magnitude relation, 
%although it was noted that either evolution away from the ZAHB or 
%dependence of the ZAHB luminosity on helium abundance could
%also affect an RR Lyrae's $\rm M^{\mathrm{RR}\,}_{V}$.
%
Based on main-sequence fitting and a period-shift analysis,
\citet{leecarney99} showed that the RR Lyrae variables in the OoII 
cluster M2 are $\sim$ 0.2 mag brighter than those in the OoI 
cluster M3, despite the fact that these clusters have similar \fehc.  If their results
can be generalized to all OoII and OoI RR Lyrae stars, this would mean that
OoII stars have brighter absolute magnitudes than OoI stars, even at the same \feh.

If the luminosities of RR Lyrae stars depend
on Oosterhoff type, are there other intrinsic differences between the 
RR Lyraes in OoI and OoII type clusters?  More importantly, could such intrinsic 
differences shed light on the origin of the Oosterhoff groups?  This is addressed 
by investigating the dereddened minimum light $\rm(\vmr)$, $\rm(\vmr)_{min,0}$,
of the fundamental mode RR Lyrae stars in the OoI 
cluster IC$\,$4499 and the OoII cluster M15.  Because color is a temperature indicator, 
and because $\rm(\vmr)_{min,0}$ is independent of period, amplitude and
metallicity \citep{kunder10}, potential temperature
differences inherent in the different Oosterhoff groups would be elucidated.

\begin{figure}[htb]
\includegraphics[width=16cm]{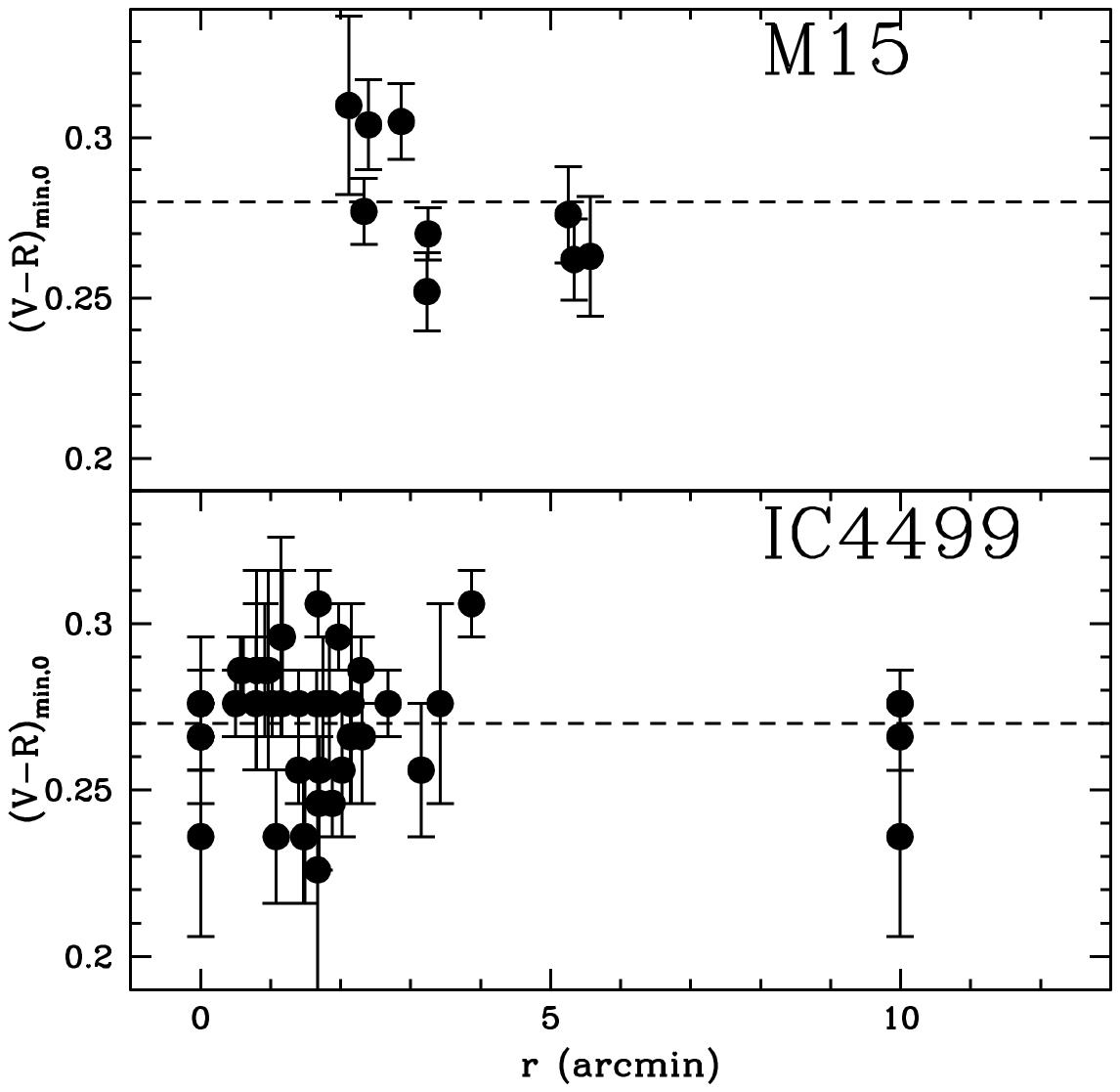}
\caption{The dereddened minimum light $\rm (\vmr)$ color of the RR Lyrae 
variables in the OoI cluster IC$\,$4499 and OoII cluster M15 is plotted 
against $r$, the distance from the cluster center.  The mean 
$\rm (\vmr)_{min,0}$ for both clusters is indicated by a dashed line and is
essentially the same as the $\rm (\vmr)_{min,0}$ of local field RR Lyrae stars.
\label{redmin}}
\end{figure}

Figure~\ref{redmin} shows the dereddened minimum light color of the IC$\,$4499 
and M15 RR0 variables as a function of radial distance from the center of the cluster.  
The radial distance is used to recognize possible differential reddening effects.
The adopted color excess for M15 is $\rm E(B-V)$=0.10 \citep{harris96}, which is
also supported by analysis of deep {\it Hubble Space Telescope\/} photometry
\citep{recio05}.  The adopted color excess for IC$\,$4499 is $\rm E(B-V)$=0.23 
\citep{harris96}.  This value is between $\rm E(B-V)$=0.22, found by 
\citet{walker96} using four different estimates of reddening to the cluster,
and $\rm E(B-V)$=0.24, found by \citep{storm04} using the mean $V$- and 
$K$-band photometry of RR Lyrae variables in IC$\,$4499.  The 
minimum-light color is found by averaging the $\rm (\vmr)$ color over
the phase interval 0.5--0.8, and the $\rm(\vmr)_{min}$ colors for the 
IC$\,$4499 are presented in Table~3.  For the M15 RR Lyrae variables,
only those with light curves 
that \citet{silbermann95} consider ``good'' were used, and these stars 
have photometric standard errors of $\sim$0.015 mag in $\rm (\vmr)$.
To convert from $\rm E(B-V)$ to $\rm E(V-R)$, it is assumed that
$\rm E(V-R) = 0.77 \ E(B-V)$ from Cardelli, Clayton \& Mathis (1989).

The average dereddened $\rm (\vmr)$ at minimum light, $\rm (\vmr)_{min,0}$ for
the M15 and IC$\,$4499 RR0 variables is 0.28 $\pm$ 0.02 and
0.27 $\pm$ 0.02, respectively, where 0.02 is the dispersion about the mean.  
The standard error of the mean is 0.003 and 0.007 respectively,
although these numbers do not include any possible systematic errors 
in the photometric calibration of either study, which could---in our 
experience---be of order 0.01~mag.   Not only are these values almost 
identical to the $\rm (\vmr)_{min,0}$ for local RR Lyrae stars \citep{kunder10}, 
but these values are within the errors of each other.  Any differential 
reddening is negligible compared to the 
errors in $\rm <\vmr>_{min,0}$.  At least for this OoI and OoII cluster pair, 
no difference in the dereddened colors at minimum light is seen in RR Lyrae
variables at the 0.02 mag level. 

This result suggests that the intrinsic color differences between 
the OoI and OoII RR Lyraes are either very small, or do not occur at 
minimum light.  As $\rm (\vmr)$ color is a good indicator of temperature
($e.g.$, Barnes \& Evans 1976), our 
result also suggests that at minimum light OoI and OoII RR Lyraes
have very similar temperatures.  A 0.02 mag uncertainty translates 
into a temperature difference between the OoI and OoII RR Lyraes 
of $\sim$100 K at minimum light.  This is calculated  
from the \citet{barnes77} relationship relating effective temperature 
to dereddened Johnson $\rm (\vmr)$ from 
Cepheid pulsation, and agrees with model atmosphere parameters
appropriate to RR Lyrae stars from Manduca \& Bell 1981 (their Table~1)
for both Johnson and Cousins colors\footnote{We note that the Johnson
$R$ used in \citet{barnes77} is not the same as the Kron Cousins $R$ used
here.}.

In order to fully understand observed properties of RR Lyrae variables, it is 
necessary to study those properties as a function of phase 
\citep[see $e.g.$,][]{kanbur05}.  Although the shape of the light curve
affects the estimate of the mean magnitude and colors (Bono, Caputo
\& Stellingwef 1994) it appears that at minimum light, RR Lyraes
are remarkably similar.  It would be worthwhile to perform a more
comprehensive analysis of the various period-color and amplitude-color 
relations as functions of pulsation phase and Oosterhoff type.  This is beyond 
the scope of this paper, as detailed knowledge of how the metallicity, 
period and amplitude affect other pulsation phases would be needed.

Given the homogenous photometry and reddening of the RR Lyraes
in IC$\,$4499, it is promising to search for any $\rm (\vmr)_{min,0}$ dependence
on period or amplitude.  A weighted least squares analysis of 
$\rm (\vmr)_{min,0}$ as a function of period reveals a slope significant
at the 2 $\sigma$ level (0.09 $\pm$ 0.04 mag day$^{-1}$).  Although not 
formally significant, this may suggest $\rm (\vmr)_{min,0}$ does
have some dependence on period at the $\sim$0.01 mag level.
A weighted least squares analysis of $\rm (\vmr)_{min,0}$ as a 
function of amplitude reveals no correlation
(the slope found is $-$0.008 $\pm$ 0.045 mag mag$^{-1}$).

\section{Discussion}
A very basic prediction of stellar evolution and pulsation theory is that
the period of a star changes with time.  The
luminosity and temperature of a star will slowly change, and correspondingly
its mean density will also change.  From the \citet{ritter79} period-mean 
density relation of pulsation theory, $P\sqrt{<\rho>}$ $\sim$ const., the 
period should change inversely with the density.
The models of  \citet{lee91} clearly show a predicted trend of increasing period change
rates for bluer HB types (which are generally OoII clusters) and decreasing period 
change rates for redder HB types (generally OoI clusters).  Among the more recent papers,
we mention \citet{catelan09} and \citet{nemec04} who fit these
models to the observed GC RR Lyrae period change rates.  Although
in general the large uncertainties in the mean period change rates
do overlap these models, the chi squared values are not consistent
with a good fit ($\chi^2$$\sim$0.08).  The above analysis indicates why this
may be the case--there is a predisposition for positive period change rates
in both OoI and OoII RR Lyrae variables in the Milky Way GCs.  No cluster
of either Oosterhoff group exhibits an abundance of RR Lyrae variables with
decreasing periods.

The RR0 Lyrae variables in $\omega$ Cen and in the field confirm
this result.  Although most RR Lyrae stars in $\omega$ Cen are known to be similar to 
those in OoII clusters, a smaller population are similar to RR Lyraes in OoI 
clusters (Butler, Dickens, \& Epps 1978; Caputo 1981; Clement \& Rowe 2000).  
These two populations are easily distinguished in the PA plane;  however, there 
is no evidence of a correlation of period change rate as a function 
of their position on the period-amplitude diagram.  The same is true for the field 
RR Lyrae variables, which are also shown to be a mix of
RR Lyraes with OoI and OoII properties \citep{miceli08, delee08, szczygiel09, bono97c}.  
Assuming that the period-amplitude diagram can be effectively used to classify 
RR Lyrae stars into an Oosterhoff type, this means that evolution
and Oosterhoff type are not correlated.  One of the most promising interpretations of the
double main sequence in the CMD of $\omega$ Cen is a helium abundance variation
\citep{norris04, piotto05}, which may suggest that helium abundance variations contributes
to the Oosterhoff types.  From high-resolution spectroscopy,
\citet{sollima06} found that the helium enrichment seen in the 
$\omega$~Cen blue main sequence stars is absent in the metal intermediate 
RR Lyrae population.  Hence at least for $\omega$~Cen, two populations with 
similar metallicities but very different helium abundances seem to 
coexist within the cluster.  Further, as shown from nonlinear pulsation
models including a non-local time-dependent treatment of convection
from $e.g.$, \citet{marconi09} and \citet{bono97a}, differences in helium can 
significantly contribute to the position of an RR Lyrae in the period-amplitude diagram. 

If differences in HB evolution are the explanation for the different Oosterhoff types, 
RR Lyrae stars in OoII cluster would begin their lives at higher temperatures
and evolve to lower temperatures, exhibiting positive period change rates as they
evolve.  From the minimum light colors of the RR Lyrae variables in the 
OoI cluster IC$\,$4499 and the OoII cluster M15, no difference in temperature at minimum
light is seen to within $\sim$ 100 K.  It would be interesting to investigate any temperature
and color differences between RR Lyrae variables in different Oosterhoff type GCs as a function
of other pulsational phases.  

Recently two metal-rich GCs, NGC$\,$6388 and NGC$\,$6441 ($\rm [Fe/H]$$\sim$$-$0.6 dex), 
were found to host a number of RR Lyrae variables with properties that
do not follow the period-metallicity trends of the other two Oosterhoff groups \citep{pritzl01}.
The RR Lyrae variables in these GCs occupy the OoII region in the PA plane and have mean
periods similar to OoII GCs, but shifted to even longer periods.  However, they
have metallicities similar to, or even higher than, OoI GCs.
Termed as ``Oosterhoff III" GCs, it has been shown by \citet{pritzl00} that 
the RR Lyraes found in these GCs can not be totally accounted for by evolutionary effects.
Although these GCs are peculiar in their own right, this result is consistent 
with our findings that evolution and Oosterhoff type are not correlated.
As detailed by \citet{busso07}, a possible solution 
to the peculiar morphology of the HB in both clusters is the presence of multiple stellar
populations with different initial He contents.  \citet{caloi07} have also shown that the 
mean period of RR Lyrae in these two strange clusters 
can be reproduced by assuming a strong He enhancement.

It is also worth mentioning NGC$\,$4147, an OoI globular cluster \citep{stetson05}.  This GC has
one of the bluest HBs and is one of the most metal-poor OoI clusters.  
\citet{contreras05} point out that an OoI classification for NGC$\,$4147 is 
inconsistent with the theoretical paradigm that RR Lyrae stars 
in OoI globular clusters are relatively unevolved objects, whereas those in OoII globulars are evolved from a position on the blue ZAHB.  Further, this globular cluster, as well as the 
OoI GCs NGC$\,$6171 and M62 have a large number fraction
of RR$1$ variables to RR0 variables, which is in stark contrast to the theory that OoI 
GCs are less evolved than OoII GCs \citep{contreras05}.  

Seventeen double-mode RR Lyrae stars (RR$01$ stars) in IC$\,$4499 have been 
identified by \citet{clement86} and \citet{walker96}.  Because the period ratios
of RR$01$ stars have a strong dependence on mass, derivation of masses for
RR Lyrae stars can be obtained ($e.g.$, Cox, King \& Hodson 1980, 
Cox, Hodson \& Clancy 1983).
\citet{clement86} found that the IC$\,$4499 RR$01$ stars are similar 
to those stars in the OoI cluster M3, but considerably different from 
RR$01$ stars found in various OoII systems.  
In the period ratio diagram (used for stars that pulsate with at least two 
periods simultaneously; the ratio of the periods is
plotted as a function of longer period), the RR$01$ stars divide 
into two groups, split according to Oosterhoff type.  
Since then, \citep{clementini04} have found two M3 RR01 stars with 
period ratios so low, that they lie well separated from all previously 
known RR01 stars in the Petersen diagram. 
This large spread in period ratio of the M3 double-mode pulsators is most 
likely due to spread in mass, and \citet{clementini04} suggest this anomalous 
spread in mass could be caused by mass-transfer in binary systems, helium 
enhancement, or, less likely, varying $\alpha$-element 
enhancement among the M3 stars.  \citet{clementini04} also point out that the scenario
that OoI clusters such as M3 are evolving blueward does not fit well,  since 
both blueward and redward evolution seems to occur among the 
M3 double-mode RR Lyrae stars.

\section{Conclusions}
Period change rates for the RR Lyraes in IC$\,$4499 are calculated and it is 
found that there is a preference for the RR Lyrae variables in IC$\,$4499 to 
have positive period change rates.  These period change rates are an order
of magnitude larger than predicted by HB evolution models.  Although
a systematic error of $\pm$0.07 d Myr$^{-1}$ in $\beta$
would reconcile difference between theory and the period change observations 
presented here, the error analysis in the period change rate determination does 
not easily lend itself to such a larger error.  Recent period change rate
determinations from time baselines of 100 years \citep{leborgne07, jurcsik01}
equally present challenges to the period change rates predicted by the
HB models.  

If the period change rates can be attributed
to stellar evolution, this surplus of period increases is in
contrast to the explanation that the RR Lyraes in OoI clusters are less 
evolved than those in OoI clusters and preferentially traverse the instability strip 
from red to blue.  Comparisons with the OoI clusters NGC$\,$7006, M3 and M5 
also show a preponderance for RR Lyraes with positive period change rates,
which may imply that the RR Lyraes in the OoI clusters also do not preferentially evolve 
from red to blue along the horizontal branch. 
Comparisons of the period change rates of the OoII clusters M15, M2 and 
NGC$\,$5053, the field RR0 variables, and the mixed cluster $\omega$~Cen, 
imply this as well.  Although OoII RR0 variables
cluster along a specific plane in the $\rm PA_V$ diagram, there is no clear preference
for these RR Lyrae variables to have either positive or negative period change rates.
Hence if the period change rates are an indication of evolution, 
differences in the evolutionary states of the OoI and OoII RR Lyrae variables are
not seen.   This would strongly suggest that age is not the primary answer in explaining the 
Oosterhoff types.  

We have discussed the dereddened $\rm (\vmr)$ colors and temperatures at 
minimum light of the OoI cluster IC$\,$4499 and the OoII cluster M15.  
It is found that to within 0.02 mag, there is no difference in the minimum light 
dereddened colors of the RR0 variables from these two clusters.  
This is an especially useful property given that there are
studies showing that 
at similar \feh metallicities, the absolute magnitudes of RR Lyrae stars in 
OoI and OoII clusters can differ by $\sim$0.2 mag \citep{leecarney99, kunder09}.   
In terms of $\rm T_{eff}$, this suggests that the temperatures of the 
RR Lyraes in Oosterhoff I and Oosterhoff II globular clusters are within 
$\sim$100 K at minimum light.  If variables in OoII clusters 
are more evolved than those in OoI clusters, this temperature similarity must
be explained by a difference in mass, luminosity or Helium content among
possible effects.

\acknowledgements
This research draws upon data provided by the NOAO Science Archive. NOAO is operated by the Association of Universities for Research in Astronomy (AURA) under cooperative agreement with the National Science Foundation.  This paper includes data gathered with the 6.5 meter Magellan Telescopes located at Las Campanas Observatory, Chile.
This work has made use of BaSTI web tools.
MZ is supported by the FONDAP Center for Astrophysics 15010003, the BASAL
CATA PFB-06, Fondecyt Regular 1085278 and the MIDEPLAN Milky Way Millennium
Nucleus P07-021-F.  This project was partially supported by PRIN INAF 2010 (P.I. R. Gratton).

\clearpage

\begin{center}
\begin{scriptsize}
\begin{longtable}{cccccc}
\caption{Period Change Rate of IC$\,$4499 RR Lyrae Stars} \label{grid_mlmmh} \\
\hline \hline \\[-2ex]
 \multicolumn{1}{c}{\textbf{Number}} &
   \multicolumn{1}{c}{\textbf{Period(d)}} &
   \multicolumn{1}{c}{\textbf{$\beta$ (d Myr$^{-1}$)}} &
   \multicolumn{1}{c}{\textbf{$\sigma_\beta$}} &
   \multicolumn{1}{c}{\textbf{$\alpha$ (1/10$^{-10}$d)}} &
  \multicolumn{1}{c}{\textbf{$2\times \sigma_{\Delta P / \Delta t}$(d Myr$^{-1}$)}}  \\[0.5ex] \hline
   \\[-1.8ex]
\endfirsthead

%This is the header for the remaining page(s) of the table...
\multicolumn{6}{c}{{\tablename} \thetable{} -- Continued} \\[0.5ex]
  \hline \hline \\[-2ex]
   \multicolumn{1}{c}{\textbf{Number}} &
   \multicolumn{1}{c}{\textbf{Period (d)}} &
   \multicolumn{1}{c}{\textbf{$\beta$ (d Myr$^{-1}$)}} &
   \multicolumn{1}{c}{\textbf{$\sigma_\beta$}}  &
   \multicolumn{1}{c}{\textbf{$\alpha$ (1/10$^{-10}$d)}} &
  \multicolumn{1}{c}{\textbf{$2\times \sigma_{\Delta P / \Delta t}$(d Myr$^{-1}$)}} \\[0.5ex] \hline
  \\[-1.8ex]
\endhead

%This is the footer for all pages except the last page of the table...
  \multicolumn{4}{l}{{Continued on Next Page\ldots}} \\
\endfoot

%This is the footer for the last page of the table...
  \\[-1.8ex] \hline \hline
\endlastfoot
2 & 0.4936367 & 0.12 & 0.03 & 6.49 & 0.12 \\
3 & 0.4832464 & 0.05 & 0.02 & 2.61 & 0.07 \\
4 & 0.6236188 & 0.18 & 0.05 & 7.77 & 0.36 \\
5 & 0.5568808 & -0.30 & 0.08 & -14.90 & 0.04 \\
6 & 0.5778506 & 8.7 & 0.1 & 238.3 & 1.1 \\ 
9 & 0.7096094 & 0.01 & 0.02 & 0.016 & 0.02 \\
11 & 0.6314811 & 0.04 & 0.03 & 1.65 & 0.01 \\
12 & 0.5947952 & 0.17 & 0.06 & 8.01 & 0.03 \\
13 & 0.5115214 & -0.29 & 0.05 & -15.52 & 0.36 \\
16 & 0.5597099 & 0.10 & 0.07 & 4.65 & 0.03 \\
17 & 0.4980768 & 0.38 & 0.11 & 20.83 & 0.22 \\
24 & 0.5165604 & -0.31 & 0.08 & -16.17 & 0.07 \\
25 & 0.6021555 & -0.03 & 0.11 & -3.76 & 0.06 \\
27 & 0.5067589 & 0.52 & 0.02 & 28.310 & 0.09 \\
28 & 0.5827866 & 0.21 & 0.10 & 9.866 & 0.09 \\
29 & 0.3626859 & 0.03 & 0.04 & 2.567 & 0.08 \\
34 & 0.4935571 & 0.08 & 0.133 & 4.55 & 0.06 \\
37 & 0.5648090 & 0.33 & 0.03 & 16.19 & 0.06 \\
44 & 0.5253868 & -0.48 & 0.06 & -24.96 & 0.03 \\
46 & 0.5747031 & -3.5 & 0.4 & -95.9 & 0.13 \\
47 & 0.5832672 & 2.1 & 0.1 & 57.3 & 0.8 \\ 
49 & 0.4991130 & 1.1 & 0.1 & 29.9 & 0.7 \\ 
50 & 0.5657635 & 1.59 & 0.02 & 43.5 & 0.01 \\
51 & 0.3551330 & 0.56 & 0.16 & 42.94 & 1.61 \\ 
54 & 0.7364995 & -0.91 & 0.22 & -33.6 & 0.02 \\
55 & 0.3585305 & 0.16 & 0.03 & 12.04 & 0.92 \\
64 & 0.6361433 & -0.06 & 0.16 & -2.71 & 0.07 \\
66 & 0.6260572 & 0.22 & 0.03 & 9.62 & 0.10 \\
69 & 0.340198 & 4.25 & 0.51 & 116.3 & 2.27 \\ 
70 & 0.5370717 & 1.9 & 0.05 & 53.1& 0.32 \\
72 & 0.6813507 & 0.32 & 0.09 & 12.90 & 0.02 \\
74 & 0.6023051 & 0.95 & 0.07 & 43.18 & 0.04 \\ 
76 & 0.6416066 & 0.84 & 0.21 & 35.93 & 0.02 \\ 
77 & 0.3546419 & -0.20 & 0.01 & -15.52 & 0.09 \\
85 & 0.5939029 & 0.46 & 0.32 & -20.56 & 0.07 \\
89 & 0.3553893 & 0.12 & 0.01 & 9.24 & 0.12 \\
92 & 0.3547447 & 0.17 & 0.02 & 13.35 & 0.25 \\
95 & 0.3626458 & 0.14 & 0.02 & 10.27 & 0.23 \\
98 & 0.3556421 & 0.51 & 0.06 & 39.47 & 0.02 \\
103 & 0.3529834 & 0.11 & 0.03 & 8.76 & 1.05 \\
108 & 0.6414954 & -0.66 & 0.16 & -27.96 & 0.12 \\
111 & 0.3573436 & 0.14 & 0.01 & 11.03 & 0.02 \\
\hline
\label{taboc}
\end{longtable}
\end{scriptsize}
\end{center}

\clearpage

\begin{table}
\begin{scriptsize}
\centering
\caption{Period Change Rates as a function of position on the $\rm PA_V$ diagram}
\label{tab1}
\begin{tabular}{lccccc} \\ \hline
Cluster & Oosterhoff section & mean (d Myr$^{-1}$) & median  (d Myr$^{-1}$) & mode  (d Myr$^{-1}$) & $N_{RR0}$ \\ \hline
NGC 7006 & OoI$\rm - {PA_V}$ & 0.04 $\pm$ 0.02 & 0.03 & 0.03 & 34  \\
M3 & OoI$\rm-{PA_V}$ & 0.05 $\pm$ 0.04 & 0.11 & 0.11 & 33  \\
M15 & OoI$\rm-{PA_V}$ & $-$0.03 $\pm$ 0.04 & 0.02 & 0.01 & 13 \\
IC$\,$4499 & OoI$\rm-{PA_V}$ & 0.02 $\pm$ 0.20 & 0.17 & 0.08 & 19 \\
OoI  GCs & OoI$\rm-{PA_V}$ & 0.04 $\pm$ 0.02 & 0.03 & 0.03 & 80 \\
OoII GCs & OoI$\rm-{PA_V}$ & $-$0.08 $\pm$ 0.11 & 0.02 & 0.05 & 7 \\
$\omega~$Cen & OoI$\rm-{PA_V}$ & 0.05 $\pm$ 0.10 & 0.03 & 0.07 & 13 \\
Field & OoI$\rm-{PA_V}$ & $-$0.04 $\pm$ 0.05 & $-$0.07 & $-$0.07 & 24 \\
\hline
\hline
OoI GCs & OoII$\rm-{PA_V}$ & $-$0.09 $\pm$ 0.15 & 0.04 & 0.2 & 10 \\
OoII GCs & OoII$\rm-{PA_V}$ & 0.11 $\pm$ 0.04 & 0.08 & 0.15 & 27 \\
$\omega~$Cen & OoII$\rm-{PA_V}$ & $-$0.26 $\pm$ 0.48 & 0.13 & 0.18 & 42 \\
Field & OoII$\rm-{PA_V}$ & 0.02 $\pm$ 0.11 & 0.08 & 0.07 & 10 \\
\hline
\end{tabular}
\end{scriptsize}
\end{table}

\clearpage

\begin{table}
\begin{scriptsize}
\centering
\caption{Minimum Light Colors the IC$\,$4499 RR0 Lyrae Stars}
\label{tabmin}
\begin{tabular}{lc} \\ \hline
Star Number & $(V-R)_{min}$ \\ \hline
1 & 0.46 $\pm$ 0.01 \\
2 & 0.46 $\pm$ 0.02 \\
3 & 0.45 $\pm$ 0.02 \\
4 & 0.47 $\pm$ 0.01 \\
5 & 0.46 $\pm$ 0.01 \\
6 & 0.47 $\pm$ 0.01 \\
7 & 0.46 $\pm$ 0.03 \\
9 & 0.45 $\pm$ 0.02 \\
11 & 0.46 $\pm$ 0.01 \\
12 & 0.46 $\pm$ 0.03 \\
13 & 0.46 $\pm$ 0.02 \\
14 & 0.46 $\pm$ 0.01 \\
15 & 0.46 $\pm$ 0.01 \\
16 & 0.44 $\pm$ 0.01 \\
17 & 0.46 $\pm$ 0.01 \\
19 & 0.48 $\pm$ 0.06 \\
20 & 0.46 $\pm$ 0.01 \\
23 & 0.43 $\pm$ 0.02 \\
24 & 0.47 $\pm$ 0.03 \\
25 & 0.48 $\pm$ 0.01 \\
26 & 0.49 $\pm$ 0.01 \\
27 & 0.43 $\pm$ 0.01 \\
28 & 0.45 $\pm$ 0.01 \\
30 & 0.42 $\pm$ 0.05 \\
34 & 0.41 $\pm$ 0.05 \\
36 & 0.39 $\pm$ 0.07 \\
37 & 0.47 $\pm$ 0.03 \\
38 & 0.44 $\pm$ 0.01 \\
44 & 0.42 $\pm$ 0.02 \\
47 & 0.49 $\pm$ 0.01 \\
48 & 0.42 $\pm$ 0.02 \\
49 & 0.41 $\pm$ 0.04 \\
50 & 0.42 $\pm$ 0.03 \\
52 & 0.44 $\pm$ 0.02 \\
41 & 0.45 $\pm$ 0.02 \\
57 & 0.47 $\pm$ 0.02 \\
58 & 0.42 $\pm$ 0.02 \\
61 & 0.46 $\pm$ 0.02 \\
64 & 0.46 $\pm$ 0.02 \\
70 & 0.48 $\pm$ 0.03 \\
72 & 0.46 $\pm$ 0.01 \\
74 & 0.45 $\pm$ 0.01 \\
82 & 0.42 $\pm$ 0.03 \\
83 & 0.46 $\pm$ 0.01 \\
84 & 0.47 $\pm$ 0.02 \\
88 & 0.47 $\pm$ 0.01 \\
106 & 0.47 $\pm$ 0.01 \\
108 & 0.44 $\pm$ 0.02 \\
112 & 0.48 $\pm$ 0.02 \\
\hline
\end{tabular}
\end{scriptsize}
\end{table}

\clearpage

\clearpage

\clearpage

\clearpage

\clearpage

\end{document}